\documentclass[draftclsnofoot, onecolumn, twoside, a4paper]{IEEEtran}

\def\fI{{\mathfrak I}}
\def\fJ{{\mathfrak J}}

\def\fR{{\mathfrak R}}

\def\IE{{\mathbb E}}
\def\IF{{\mathbb F}}

\def\IN{{\mathbb N}}

\def\IR{{\mathbb R}}

\def\IZ{{\mathbb Z}}

\def\I{{\mathcal I}}

\def\R{{\mathcal R}}
\def\S{{\mathcal S}}
\def\T{{\mathcal T}}

\def\sK{{\mathscr K}}

\def\sM{{\mathscr M}}

\def\sP{{\mathscr P}}

\def\sS{{\mathscr S}}

\def\sX{{\mathscr X}}
\def\sY{{\mathscr Y}}
\def\sZ{{\mathscr Z}}

\def\bA{{\mathbf A}}

\def\bP{{\mathbf P}}
\def\bQ{{\mathbf Q}}
\def\bR{{\mathbf R}}
\def\bS{{\mathbf S}}

\def\bU{{\mathbf U}}

\def\bX{{\mathbf X}}

\def\bx{{\mathbf x}}
\def\by{{\mathbf y}}
\def\bz{{\mathbf z}}

\newcommand{\abs}[1]{\left\lvert #1 \right\rvert}
\newcommand{\Prb}[1]{\Pr\left\{ #1 \right\}}
\newcommand{\cPrb}[2]{\Pr\left\{ \left. #1 \right| #2 \right\}}
\newcommand{\diag}[1]{\textup{diag}\left\{ #1 \right\}}

\usepackage{amsfonts, amsmath, amsthm, amstext, amssymb, mathrsfs, graphicx, color, url, multirow, tabularx}
\usepackage[colorlinks=false, linkcolor=blue]{hyperref}

\newtheorem{lemma}{Lemma}[section]
\newtheorem{theorem}[lemma]{Theorem}
\newtheorem{proposition}[lemma]{Proposition}
\newtheorem{corollary}[lemma]{Corollary}
\newtheorem{example}[lemma]{Example}

\newtheorem{definition}[lemma]{Definition}

\newtheorem{remark}{Remark}
\newtheorem{problem}{Problem}

\everymath{\displaystyle}

\ifCLASSINFOpdf
\else
\fi

\begin{document}

\def\TITLE{{Coding for Computing Irreducible Markovian Functions of Sources with Memory}}
\title{\TITLE}

\author{Sheng~Huang,~\IEEEmembership{Student~Member,~IEEE,}
        Mikael~Skoglund,~\IEEEmembership{Senior~Member,~IEEE}
\thanks{S. Huang and M. Skoglund are with the Communication Theory Lab,
School of Electrical Engineering, KTH Royal Institute of Technology, Stockholm,
10044, Sweden e-mail: (sheng.huang@ee.kth.se; skoglund@ee.kth.se).}
\thanks{This work was funded in part by the Swedish Research Council.}}

\markboth{SUBMITTED}
{HUANG and SKOGLUND: \TITLE}
%

\maketitle

\begin{abstract}
  One open problem in source coding is to characterize the limits of
  representing losslessly a non-identity discrete function of the data
  encoded independently by the encoders of several correlated sources
  with
  memory. 
  This paper investigates this problem under Markovian conditions,
  namely either the sources or the functions considered are Markovian.
  We propose using linear mappings over finite rings as encoders. If
  the function considered admits certain polynomial structure, the
  linear encoders can make use of this structure to establish
  ``implicit collaboration'' and boost the performance. In fact, this
  approach universally applies to any scenario (arbitrary function)
  because any discrete function admits a polynomial presentation of
  required format.

  There are several useful discoveries in the paper. The first says
  that linear encoder over non-field ring can be equally optimal for
  compressing data generated by an irreducible Markov source.
  Secondly, regarding the previous function-encoding problem, there
  are infinitely many circumstances where linear encoder over
  non-field ring strictly outperforms its field counterpart. To be
  more precise, it is seen that the set of coding rates achieved by
  linear encoder over certain non-field rings is strictly larger than
  the one achieved by the field version, regardless which finite field
  is considered. Therefore, in this sense, linear coding over finite
  field is not optimal. In addition, for certain scenarios where the
  sources do not possess the ergodic property, our ring approach is
  still able to offer a solution.
\end{abstract}


%
\IEEEpeerreviewmaketitle

\section{Introduction}

\IEEEPARstart{T}{his} paper considers the problem of encoding a \emph{discrete function} of correlated sources with memory:

\setcounter{problem}{1}
\begin{problem}[Source Coding for Computing a Function of Sources with or without Memory]\label{prb:2}\rm
Let $S_{t}$ ($t \in \S = \{ 1, 2, \cdots, s \}$) be a \emph{source} that randomly generates discrete data
$$\cdots, X_{t}^{(1)}, X_{t}^{(2)}, \cdots, X_{t}^{(n)}, \cdots,$$
where $X_{t}^{(n)}$ has a finite sample space $\sX_{t}$ for all $n \in \IN^{+}$.
Given a discrete function $g : \sX \rightarrow \sY$, where $\sX = \prod_{t \in \S} \sX_{t}$, 
what is the biggest region $\R[g] \subset \IR^{s}$ satisfying, $\forall \; (R_{1}, R_{2}, \cdots, R_{s}) \in \R[g]$ and $\forall \; \epsilon > 0$, $\exists \; N_{0} \in \IN^{+}$, such that, $\forall \; n > N_{0}$, there exist $s$ \emph{encoders}
$
\phi_{t} : \sX_{t}^{n} \rightarrow \left[ 1, 2^{n R_{t}} \right], t \in \S,
$
and one \emph{decoder}
$
\psi : \prod_{t \in \S} \left[ 1, 2^{n R_{t}} \right] \rightarrow \sY^{n}
$
with
$$
\Pr \left\{ \vec{g}\left( X_{1}^{n}, \cdots, X_{s}^{n} \right) \neq \psi\left[ \phi_{1}\left( X_{1}^{n} \right), \cdots, \phi_{s}\left( X_{s}^{n} \right) \right] \right\} < \epsilon,
$$
where
\begin{align*}
X_{t}^{n} = & \left[ X_{t}^{(1)}, X_{t}^{(2)}, \cdots, X_{t}^{(n)} \right] \mbox{ and}\\
\vec{g}\left( X_{1}^{n}, \cdots, X_{s}^{n} \right) = & \left[ Y^{(1)}, Y^{(2)}, \cdots, Y^{(n)} \right]^{t}
\end{align*}
with $Y^{(n)} = g\left( X_{1}^{(n)}, X_{2}^{(n)}, \cdots, X_{s}^{(n)} \right)$?
\end{problem}

\noindent The region $\R[g]$ is called the \emph{achievable coding
  rate region} for computing $g$. A rate touple $\bR \in \IR^{s}$ is
said to be \emph{achievable} for computing $g$ (or simply achievable)
if and only if $\bR \in \R[g]$. A region $\R \subset \IR^{s}$ is said
to be \emph{achievable} for computing $g$ (or simply achievable) if
and only if $\R \subseteq \R[g]$.

Problem \ref{prb:2} is a generalization of \cite[Problem
1]{Huang2012d} which considers only the special case that the process
$$\cdots, X^{(1)}, X^{(2)}, \cdots, X^{(n)}, \cdots,$$
where $X^{(n)} = \left[ X_{1}^{(n)}, X_{2}^{(n)}, \cdots, X_{s}^{(n)} \right]$, in Problem \ref{prb:2} is i.i.d., so is 
$$\cdots, Y^{(1)}, Y^{(2)}, \cdots, Y^{(n)}, \cdots.$$
Related work for this special scenario includes: \cite{Slepian1973,
  Huang2013a} which considers the case that $g$ is an identity
function; \cite{Korner1979, Ahlswede1983} where $g$ is the binary
sum; \cite{Han1987, Huang2012a} for conditions under which that
$\R[g]$ is strictly larger than the Slepian--Wolf region;
\cite{Sefidgaran2011, Huang2012b, Huang2012c, Huang2013b, Huang2012d}
for an arbitrary discrete function
$g$. 
Generally speaking, $\R[g]$ is unknown in cases where $g$ is not an
identity function, and it is larger (strictly in many cases) than the
Slepian--Wolf region.

Furthermore, much less is known in the case of sources with
memory. Let
\begin{align}
\R_{s} = \bigg\{ [ R_{1}, R_{2}, \cdots, R_{s} ] \in \IR^{s} \bigg| \sum_{t \in T} R_{t} > \lim_{n \rightarrow \infty} \dfrac{1}{n} \Big[ & H\left( X^{(n)}, X^{(n-1)}, \cdots, X^{(1)} \right) \nonumber\\
 & - H\left( X_{T^{c}}^{(n)}, X_{T^{c}}^{(n-1)}, \cdots, X_{T^{c}}^{(1)} \right) \Big], \emptyset \neq T \subseteq \S \bigg\}\footnotemark, \label{eq:cover_ergodic_sources}
\end{align}\footnotetext{Assume the limits exist.}
where $T^{c} = \S \setminus T$ and $X_{T}^{(n)}$ is the random variable array $\prod_{t \in T} X_{t}^{(n)}$.
By \cite{Cover1975}, if the process
$$\cdots, X^{(1)}, X^{(2)}, \cdots, X^{(n)}, \cdots$$
is \emph{jointly ergodic} (see \cite{Cover1975} for definition), then
$\R_{s} = \R[g]$ for an identity function $g$. Naturally, $\R_{s}$ is
an inner bound of $\R[g]$ for an arbitrary $g$. However, $\R_{s}$ is
not always tight (optimal), i.e. $\R_{s} \subsetneq \R[g]$, as we will
demonstrate later in Example \ref{eg:computing_Mar_func}. Even for the
special scenario of correlated i.i.d. sources, i.e.
$$\cdots, X^{(1)}, X^{(2)}, \cdots, X^{(n)}, \cdots$$
is i.i.d., $\R_{s}$, which is then the Slepian--Wolf region, is not tight (optimal) in general as mentioned before.
Unfortunately, little is mentioned in existing literature regarding the case
$$\cdots, X^{(1)}, X^{(2)}, \cdots, X^{(n)}, \cdots$$
is not memoryless, neither for the case that
$$\cdots, Y^{(1)}, Y^{(2)}, \cdots, Y^{(n)}, \cdots$$
is Markovian (which does not necessary imply that $\cdots, X^{(1)}, X^{(2)}, \cdots, X^{(n)}, \cdots$ is jointly ergodic or Markov).

This paper focuses on Problem \ref{prb:2} in the sense that some
additional Markovian constraints are imposed since the original
scenario is too general. We assume that:
\begin{enumerate}
\item[(c1)]
There exist some finite ring $\fR$, functions $k_{t} : \sX_{t} \rightarrow \fR$ ($t \in \S$) and $h : \fR \rightarrow \sY$ with
\begin{align}
g(x_{1}, x_{2}, \cdots, x_{s}) = h \left( \sum_{t \in \S} k_{t}(x_{t}) \right), \label{eq:c1}
\end{align}
such that $\left\{ \sum_{t \in \S} k_{t}\left( X_{t}^{(n)} \right) \right\}_{-\infty}^{\infty}$ is irreducible\footnote{Irreducibility of a Markov chain / process is sometimes (implicitly) assumed in some literature.} Markovian\footnote{For any finite discrete function $g$, such a finite ring $\fR$ and functions $k_{t}$'s and $h$ always exist by Lemma \ref{lma:pre_poly_func_presentation}. However, the Markovian condition is not guaranteed in general.}.
\end{enumerate}
By Lemma \ref{lma:pre_poly_func_presentation} and Lemma \ref{lma:Markovian_func}, (c1) includes a very interesting scenario:
\begin{enumerate}
\item[(c0)]
$g$ is arbitrary, while
$$\cdots, X^{(1)}, X^{(2)}, \cdots, X^{(n)}, \cdots$$
forms an irreducible Markov chain with transition matrix
\begin{align}
\bP_{0} = c_{1} \bU + ( 1 - c_{1} ) \mathbf{1}, \label{eq:Markovian_func_cond}
\end{align}
where all rows of $\bU$ are identical to some unitary vector $[ u_{x} ]_{x \in \sX}$, $\mathbf{1}$ is an identity matrix and $0 \leq c_{1} \leq 1$.
\end{enumerate}
If, as a special case, $c_{1} = 1$, then Problem \ref{prb:2} renders to \cite[Problem 1]{Huang2012d}, since
$$\cdots, X^{(1)}, X^{(2)}, \cdots, X^{(n)}, \cdots$$
becomes i.i.d..
Actually, (c0) is very interesting because of the fact:
\begin{enumerate}
\item[]
A stationary finite-state Markov chain
$$\cdots, X^{(1)}, X^{(2)}, \cdots, X^{(n)}, \cdots$$
admits a transition matrix of the form \eqref{eq:Markovian_func_cond}, if and only if
$$\cdots, \Gamma\left( X^{(1)} \right), \Gamma\left( X^{(2)} \right), \cdots, \Gamma\left( X^{(n)} \right), \cdots$$
is Markovian for all feasible mappings $\Gamma$ \cite[Theorem 3]{Burke1958}.
\end{enumerate}
We will explain the mechanism that (c0) illustrates when the
discussion comes. Here we would like to point out that (c1)
is a rather general assumption. It even includes some scenario that
$$\cdots, X^{(1)}, X^{(2)}, \cdots, X^{(n)}, \cdots$$
does not possess the ergodic property (see Example
\ref{eg:computing_Mar_func:ext}). Therefore, \cite{Cover1975} does not
apply and \eqref{eq:cover_ergodic_sources} does not present an
achievable region. However, it is sometimes possible to classify such
a scenario as a special case of (c1), to which a solution is provided in this
paper (see Section \ref{sec:computing_Mar_func}).

This paper aims at developing similar results as \cite{Huang2012d}
based on this new setting. To be more precise, we will first prove an
achievability theorem for source coding with linear encoder over
finite ring for compressing a single finite-state irreducible Markov
source. This generalizes the corresponding theorem regarding linear
encoder over field.  Making use of the linear coding technique
introduced by this achievability theorem, we then address Problem
\ref{prb:2} of computing $g$ regarding each of the previous
conditions, (c0) and (c1). Inner bounds of $\R[g]$ are presented. It
is demonstrated that the achievable regions given by these inner
bounds are beyond \eqref{eq:cover_ergodic_sources}.  Even more
interestingly, our method (for computing some $g$) even works for
cases in which
$$\cdots, X^{(1)}, X^{(2)}, \cdots, X^{(n)}, \cdots$$
does not possess the ergodic property.  Finally, a comparison between
linear encoder over non-field ring and its field counterpart is
carried out. It is seen that the non-field ring version offers many
advantages, including strictly outperforming the field version in
terms of achieving larger achievable region for computing (infinitely)
many functions. In this sense, we conclude that linear coding over
finite field is not optimal.

Apart from classic information theoretical techniques, the key
mathematical tools involved are the uncoupling-coupling technique and
the concept of stochastic complement of finite-state Markov processes
(see \cite{Meyer1989} for more details). With the aid of these tools,
we will introduce the concept of Supremus typical sequences
(Definition \ref{def:Sup_tp}) and prove related asymptotic
properties (Proposition \ref{prop:aep_of_Sup_tp}) and typicality
lemmas (Appendix \ref{app:d}). These serve as the foundation of our
arguments thereafter.

\section{Preliminaries}

Required concepts and properties are listed in this section to
partially make the paper self-contained, at the same time, to clarify
delicate aspects of concepts and (implicit) assumptions sometimes
defined slightly differently in other literature. Readers are
recommended to go thought (quickly) to identify our notation and
universal assumptions.

\subsection{Some Notation}

Let $\sX$, $\sY$ and $\sZ$ be three countable sets with or without \emph{orders} defined, e.g.
$$
\sX = \left\{ (0, 0), (0, 1), (1, 1), (1, 0) \right\} \mbox{ and } \sY = \left\{ \alpha, \beta \right\} \times \IN^{+}.
$$
In many places hereafter, we write $[ p_{i, j} ]_{i \in \sX, j \in \sY}$ ($[ p_{i} ]_{i \in \sX}$) for a ``matrix'' (``vector'') whose ``$(i, j)$th'' (``$i$th'') entry is $p_{i, j} \; (p_{i}) \in \IR$. Matrices $\left[ p_{i, j}' \right]_{i \in \sX, j \in \sY}$ and $[ q_{j, k} ]_{j \in \sY, k \in \sZ}$ are similarly defined. Let $\bP = [ p_{i, j} ]_{i \in \sX, j \in \sY}$. For subsets $A \subseteq \sX$ and $B \subseteq \sY$, $\bP_{A, B}$ is designated for the ``submatrix'' $[ p_{i, j} ]_{i \in A, j \in B}$. We will use ``index oriented'' operations, namely
\begin{align*}
[ p_{i} ]_{i \in \sX} [ p_{i, j} ]_{i \in \sX, j \in \sY} = & \left[ \sum_{i \in \sX} p_{i} p_{i, j} \right]_{j \in \sY};\\
[ p_{i, j} ]_{i \in \sX, j \in \sY} + \left[ p_{i, j}' \right]_{i \in \sX, j \in \sY} = & \left[ p_{i, j} + p_{i, j}' \right]_{i \in \sX, j \in \sY};\\
[ p_{i, j} ]_{i \in \sX, j \in \sY} [ q_{j, k} ]_{j \in \sY, k \in \sZ} = & \left[ \sum_{j \in \sY} p_{i, j} q_{j, k} \right]_{i \in \sX, k \in \sZ}.
\end{align*}
In addition, a matrix $\bP_{A, A} = [ p_{i, j } ]_{i, j \in A}$ is said to be an \emph{identity matrix} if and only if $p_{i, j} = \delta_{i, j}$ (Kronecker delta), $ \forall \; i, j \in A$. We often indicate an identity matrix with $\mathbf{1}$ whose size is known from the context, while designate $\mathbf{0}$ as the \emph{zero matrix} (all of whose entries are $0$) of size known from the context. For any matrix $\bP_{A, A}$, its \emph{inverse} (if exists) is some matrix $\bQ_{A, A}$ such that $\bQ_{A, A} \bP_{A, A} = \bP_{A, A} \bQ_{A, A} = \mathbf{1}$.
Let $[ p_{i} ]_{i \in \sX}$ be non-negative and \emph{unitary}, i.e. $\sum_{i \in \sX} p_{i} = 1$, and $[ p_{i, j} ]_{i \in \sX, j \in \sY}$ be non-negative and $\sum_{j \in \sY} p_{i, j} = 1$ (such a matrix is termed a \emph{stochastic matrix}).
For discrete random variables $X$ and $Y$ with sample spaces $\sX$ and $\sY$, respectively, $X \sim [ p_{i} ]_{i \in \sX}$ and $(X, Y) \sim [ p_{i} ]_{i \in \sX} [ p_{i, j} ]_{i \in \sX, j \in \sY}$ state for
\begin{align*}
\Prb{X = i} = p_{i} \mbox{ and }
\Prb{X = i, Y = j} = p_{i} p_{i, j},
\end{align*}
for all $i \in \sX$ and $j \in \sY$, respectively.

\subsection{Markov Chains and Strongly Markov Typical Sequences}

\begin{definition}\rm
A (discrete) \emph{Markov chain} is defined to be a discrete stochastic process $\sM = \left\{ X^{(n)} \right\}_{-\infty}^{\infty}$ with \emph{state space} $\sX$ such that, $\forall \; n \in \IN^{+}$,
\begin{align*}
\cPrb{ X^{(n+1)} }{ X^{(n)}, X^{(n-1)}, \cdots, X^{(1)} } = \cPrb{ X^{(n+1)} }{ X^{(n)} }.
\end{align*}
$\sM$ is said to be \emph{finite-state} if $\sX$ is finite.
\end{definition}

\begin{definition}\rm
A Markov chain $\sM = \left\{ X^{(n)} \right\}_{-\infty}^{\infty}$ is said to be \emph{homogeneous} (\emph{time homogeneous}) if and only if
$$
\cPrb{ X^{(n+1)} }{ X^{(n)} } = \cPrb{ X^{(2)} }{ X^{(1)} }, \forall \; n \in \IN^{+}.
$$
\end{definition}
%

If not specified, we assume finite-state and homogeneous of all Markov chains considered throughout this paper. However, they are not necessarily \emph{stationary} \cite[pp. 71]{Cover2006}, or their \emph{initial distribution} is unknown.

\begin{definition}\rm
Given a Markov chain $\sM = \left\{ X^{(n)} \right\}_{-\infty}^{\infty}$ with a countable state space $\sX$, the \emph{transition matrix} of $\sM$ is defined to be the stochastic matrix $\bP = [p_{i, j}]_{i, j \in \sX}$, where $p_{i, j} = \cPrb{X^{(2)} = j}{X^{(1)} = i}$.
Moreover, $\sM$ is said to be \emph{irreducible} if and only if $\bP$ is \emph{irreducible}, namely, there exists no $\emptyset \neq A \subsetneq \sX$ such that $\bP_{A, A^{c}} = \bold{0}$.
\end{definition}

\begin{definition}\rm
A state $j$ of a Markov chain $\sM = \left\{ X^{(n)} \right\}_{-\infty}^{\infty}$ is said to be \emph{recurrent} if
\begin{align*}
\cPrb{ T < \infty }{ X^{(0)} = j } = 1,
\end{align*}
where $T = \inf\{ n > 0 | X^{(n)} = j \}$.
If in addition the conditional expectation
\begin{align*}
\IE\{ T | X^{(0)} = j \} < \infty,
\end{align*}
then $j$ is said to be \emph{positive recurrent}.
$\sM$ is said to be \emph{positive recurrent} if all states are positive recurrent.
\end{definition}

\begin{theorem}[Theorem 1.7.7 of \cite{Norris1998}]
  An irreducible Markov chain $\sM$ with a countable state space $\sX$
  is positive recurrent, if and only if it admits a non-negative
  unitary vector $\pi = \left[ p_{j} \right]_{j \in \sX}$, such that
  $\pi \bP = \pi$, where $\bP$ is the transition matrix of
  $\sM$. Moreover, $\pi$ is unique and is called the \emph{invariant
    (stationary) distribution}.
\end{theorem}

\begin{theorem}[Theorem 2.31 of \cite{Breuer2005}]
A finite-state irreducible Markov chain is positive recurrent.
\end{theorem}

Clearly, all irreducible Markov chains considered in this paper admit
a unique invariant distribution, since they are assumed to be
simultaneously finite-state and homogeneous (unless otherwise
specified).
%
%

\begin{definition}[Strong Markov Typicality]\rm
Let $\sM = \left\{ X^{(n)} \right\}_{-\infty}^{\infty}$ be an irreducible Markov chain with state space $\sX$, and $\bP = [ p_{i, j} ]_{i, j \in \sX}$ and $\pi = \left[ p_{j} \right]_{j \in \sX}$ be its transition matrix and invariant distribution, respectively.
For any $\epsilon > 0$, a sequence $\bx \in \sX^{n}$ of \emph{length} $n$ ($\geq 2$) is said to be \emph{strongly Markov $\epsilon$-typical} with respect to $\bP$ if
\begin{align}
& \begin{aligned}\label{eq:1:sup_tp}
\begin{cases}
\abs{ \dfrac{N(i, j; \bx)}{N(i; \bx)} - p_{i, j} } < \epsilon;\\
\abs{ \dfrac{N(i; \bx)}{n} - p_{i} } < \epsilon,
\end{cases}
\forall \; i, j \in \sX,
\end{aligned}\\
\mbox{or} &
\begin{aligned}\label{eq:2:sup_tp}
\begin{cases}
\sum_{i, j \in \sX} \abs{ \dfrac{N(i, j; \bx)}{N(i; \bx)} - p_{i, j} } < \epsilon;\\
\sum_{i \in \sX} \abs{ \dfrac{N(i; \bx)}{n} - p_{i} } < \epsilon,
\end{cases}
\end{aligned}
\end{align}
where $N(i, j; \bx)$ is the occurrences of sub-sequence $[i, j]$ in $\bx$ and $N(i; \bx) = \sum_{j \in \sX} N(i, j; \bx)$.
The set of all strongly Markov $\epsilon$-typical sequences with respect to $\bP$ in $\sX^{n}$ is denoted by $\T_{\epsilon}(n, \bP)$ or $\T_{\epsilon}$ for simplicity.
\end{definition}

\begin{remark}\rm
\eqref{eq:1:sup_tp} and \eqref{eq:2:sup_tp} is equivalent (in illustrating the asymptotic behavior of $\sM$) to
\begin{align*}
 & \abs{ \dfrac{N(i, j; \bx)}{n} - p_{i} p_{i, j} } < c \epsilon, \forall \; i, j \in \sX,\\
\mbox{and} & \sum_{i, j \in \sX} \abs{ \dfrac{N(i, j; \bx)}{n} - p_{i} p_{i, j} } < c \epsilon,
\end{align*}
for some fixed finite constant $c$, respectively.
\end{remark}

Let $\bP$ and $\pi$ be some stochastic matrix and non-negative unitary
vector. We define $H(\pi)$ and $H(\bP | \pi)$ to be $H(X)$ and $H(Y |
X)$, respectively, for jointly discrete random variables $(X, Y)$ such
that $X \sim \pi$ and $(X, Y) \sim \pi \bP$.

\begin{proposition}[AEP of Strongly Markov Typicality\footnote{Similar statements in many literature assume that the Markov chain is stationary. It is easy to generalize to irreducible Markov chain. To be rigorous, we include a proof in Appendix \ref{app:a}.}]\label{prop:aep_of_Mar_tp}
Let $\sM = \left\{ X^{(n)} \right\}_{-\infty}^{\infty}$ be an irreducible Markov chain with state space $\sX$, and $\bP = [ p_{i, j} ]_{i, j \in \sX}$ and $\pi = \left[ p_{j} \right]_{j \in \sX}$ be its transition matrix and invariant distribution, respectively.
For any $\eta > 0$, there exist $\epsilon_{0} > 0$ and $N_{0} \in \IN^{+}$, such that, $\forall \; \epsilon_{0} > \epsilon > 0$, $\forall \; n > N_{0}$ and $\forall \; \bx = \left[ x^{(1)}, x^{(2)}, \cdots, x^{(n)} \right] \in \T_{\epsilon}(n, \bP)$,
\begin{enumerate}
\item $\exp_{2}\left[ - n \left( H(\bP | \pi) + \eta \right) \right] < \Prb{ \left[ X^{(1)}, X^{(2)}, \cdots, X^{(n)} \right] = \bx } < \exp_{2}\left[ - n \left( H(\bP | \pi) - \eta \right) \right]$;

\item $\Prb{ \bX \notin \T_{\epsilon}(n, \bP) } < \eta$, where $\bX = \left[ X^{(1)}, X^{(2)}, \cdots, X^{(n)} \right]$; and

\item $\abs{ \T_{\epsilon}(n, \bP) } < \exp_{2}\left[ n \left( H(\bP | \pi) + \eta \right) \right]$.
\end{enumerate}
\end{proposition}

\begin{IEEEproof}
See Appendix \ref{app:a}.
\end{IEEEproof}

\begin{remark}\rm
For a strongly Markov $\epsilon$-typical sequence $(\bx, \by)^{t} \in \sX^{n} \times \sY^{n}$, it is not necessary that $\bx$ or $\by$ is strongly Markov $\epsilon$-typical. As a matter of fact, given an irreducible Markov chain $\left\{ \left( X^{(n)}, Y^{(n)} \right)^{t} \right\}_{-\infty}^{\infty}$, stochastic processes $\left\{ X^{(n)} \right\}_{-\infty}^{\infty}$ or $\left\{ Y^{(n)} \right\}_{-\infty}^{\infty}$ is not necessary Markov.
\end{remark}

\subsection{Rings, Ideals and Linear Mappings}

\begin{definition}\label{def:ring}\rm
The touple $[\fR, +, \cdot]$ is called a \emph{ring} if the following criteria are met:
\begin{enumerate}
\item $[\fR, +]$ is an \emph{Abelian group};

\item There exists a \emph{multiplicative identity}
$1 \in \fR$, namely, $1 \cdot a = a \cdot 1 = a$, $\forall \; a \in \fR$;

\item $\forall \; a, b, c \in \fR$, $a \cdot b \in \fR$ and $(a \cdot b) \cdot c = a \cdot (b \cdot c)$;

\item $\forall \; a, b, c \in \fR$, $a \cdot (b + c) = (a \cdot b) + (a \cdot c)$ and $(b + c) \cdot a = (b \cdot a) + (c \cdot a)$.

\end{enumerate}
\end{definition}

We often write $\fR$ for $[\fR, +, \cdot]$ when the \emph{operations} considered are known from the context. The operation ``$\cdot$'' is usually written by juxtaposition, $a b$ for $a \cdot b$, for all $a, b \in \fR$.

A ring $[\fR, +, \cdot]$ is said to be \emph{commutative} if $\forall \; a, b \in \fR$, $a \cdot b = b \cdot a$. In Definition \ref{def:ring}, the identity of the group $[\fR, +]$, denoted by $0$, is called the \emph{zero}. A ring $[\fR, +, \cdot]$ is said to be \emph{finite} if the cardinality $|\fR|$ is finite, and $|\fR|$ is called the \emph{order} of $\fR$. The set $\IZ_{q}$ of integers modulo $q$ is a commutative finite ring with respect to the \emph{modular arithmetic}.

\begin{definition}[c.f. \cite{Dummit2003}]\label{def:characteristic}\rm
The \emph{characteristic} of a finite ring $\fR$ is defined to be the smallest positive integer $m$, such that $\sum_{j=1}^{m} 1 = 0$, where $0$ and $1$ are the zero and the multiplicative identity of $\fR$, respectively. The characteristic of $\fR$ is often denoted by $\textup{Char}(\fR)$.
\end{definition}

\begin{remark}\rm
Clearly, $\textup{Char}(\IZ_{q}) = q$. For a finite field $\IF$, $\textup{Char}(\IF)$ is always the prime $q_{0}$ such that $\abs{ \IF } = q_{0}^{n}$ for some integer $n$ \cite[Proposition 2.137]{Rotman2010}.
\end{remark}

\begin{definition}\label{def:ideal}\rm
A subset $\fI$ of a ring $[\fR, +, \cdot]$ is said to be a \emph{left ideal} of $\fR$, denoted by $\fI \leq_{l} \fR$, if and only if
\begin{enumerate}
\item $[\fI, +]$ is a subgroup of $[\fR, +]$;

\item\label{itm:ideal2} $\forall \; x \in \fI$ and $\forall \; r \in \fR$, $r \cdot x \in \fI$.

\end{enumerate}
If condition \ref{itm:ideal2}) is replaced by
\begin{enumerate}
\item[3)] $\forall \; x \in \fI$ and $\forall \; r \in \fR$, $x \cdot r \in \fI$,

\end{enumerate}
then $\fI$ is called a \emph{right ideal} of $\fR$, denoted by $\fI \leq_{r} \fR$.
$\{0\}$ is a \emph{trivial} left (right) ideal, usually denoted by $0$.
\end{definition}

It is well-known that if $\fI \leq_{l} \fR$ or $\fI \leq_{r} \fR$, then $\fR$ is divided into disjoint \emph{cosets} which are of equal size (cardinality). $\abs{\fI}$ is called the \emph{order} of $\fI$ if it is finite. For any coset $\fJ$, $\fJ = x + \fI = \left\{ x + y | y \in \fI \right\}$, $\forall \; x \in \fJ$. The set of all cosets forms a \emph{quotient group}, denoted by $\fR / \fI$ (see \cite[Ch. 1.6 and Ch. 2.9]{Rotman2010} for more details).

\begin{definition}\label{def:linear_mapping}\rm
A mapping $f : \fR^{n} \rightarrow \fR^{m}$ given as:
\begin{align*}
f(x_{1}, x_{2}, \cdots, x_{n}) = & \big( \textstyle\sum_{j=1}^{n} a_{1, j} x_{j}, \cdots, \sum_{j=1}^{n} a_{m, j} x_{j} \big)^{t}\\
\Big( f(x_{1}, x_{2}, \cdots, x_{n}) = & \big( \textstyle\sum_{j=1}^{n} x_{j} a_{1, j}, \cdots, \sum_{j=1}^{n} x_{j} a_{m, j} \big)^{t} \Big), \nonumber\\
 & \forall \; (x_{1}, x_{2}, \cdots, x_{n}) \in \fR^{n},
\end{align*}
where $a_{i, j} \in \fR$ for all feasible $i$ and $j$, is called a \emph{left} (\emph{right}) \emph{linear mapping} over ring $\fR$. If $m = 1$, then $f$ is called a \emph{left} (\emph{right}) \emph{linear function} over $\fR$. The matrix $\bA = [a_{i, j}]_{1 \leq i, j \leq n}$ is called the \emph{coefficient matrix} of $f$.
\end{definition}

In our later discussions, we mainly use left linear mappings (functions, encoders). They are simply referred to as linear mappings (functions, encoders). This will not give rise to confusion because left linearity and right linearity can always be distinguished from the context.

\subsection{Polynomial Functions}

\begin{definition}\rm
A \emph{polynomial function} of $k$ \emph{variables} over a finite ring $\fR$ is a function $g : \fR^{k} \rightarrow \fR$ of the form
\begin{align}
g(x_{1}, x_{2}, \cdots, x_{k}) = \sum_{j=0}^{m} a_{j} x_{1}^{m_{1 j}} x_{2}^{m_{2 j}} \cdots x_{k}^{m_{k j}}, \label{eq:polynomial}
\end{align}
where $a_{j} \in \fR$ and $m$ and $m_{i j}$'s are non-negative integers.
The set of all the polynomial functions of $k$ variables over ring $\fR$ is designated by $\fR[k]$.
\end{definition}

\begin{remark}\rm
\emph{Polynomial} and polynomial function are sometimes only defined over a commutative ring \cite{Rotman2010}. It is a very delicate matter to define them over a non-commutative ring \cite{Hungerford1980, Lam2001}, due to the fact that $x_{1} x_{2}$ and $x_{2} x_{1}$ can become different objects. We choose to define ``polynomial functions'' with formula \eqref{eq:polynomial} because those functions are within the scope of this paper's interest.
\end{remark}

\begin{lemma}\label{lma:polynomial}
For any discrete function
$
g : \prod_{i=1}^{k} \sX_{i} \rightarrow \sY
$
with $\sX_{i}$'s and $\sY$ being finite,
there always exist a finite ring (field) and a polynomial function $\hat{g} \in \fR[k]$ such that
$$
\nu\left( g\left( x_{1}, x_{2}, \cdots, x_{k} \right) \right) = \hat{g}\left( \mu_{1}(x_{1}), \mu_{2}(x_{2}), \cdots, \mu_{k}(x_{k}) \right)
$$
for some injections $\mu_{i} : \sX_{i} \rightarrow \fR$ ($1 \leq i \leq k$) and $\nu :  \sY \rightarrow \fR$.
\end{lemma}

\begin{IEEEproof}
Let $p$ be a prime such that $p^{m} \geq \max\left\{ \abs{\sY}, \abs{\sX_{i}} \left| 1 \leq i \leq k \right. \right\}$ for some integer $m$, and choose $\fR$ to be a finite field of order $p^{m}$.
By \cite[Lemma 7.40]{Lidl1997}, the number of polynomial functions in $\fR[k]$ is $p^{m p^{m k}}$. Moreover, the number of distinct functions with domain $\fR^{k}$ and codomain $\fR$ is also $\abs{\fR}^{ \abs{\fR^{k}} } = p^{m p^{m k}}$. Hence, any function $g : \fR^{k} \rightarrow \fR$ is a polynomial function.

In the meanwhile, any injections $\mu_{i} : \sX_{i} \rightarrow \fR$
($1 \leq i \leq k$) and $\nu : \sY \rightarrow \fR$ give rise to a
function
\begin{align*}
\hat{g} = \nu \circ g\left( \mu_{1}', \mu_{2}', \cdots, \mu_{k}' \right) : \fR^{k} \rightarrow \fR,
\end{align*}
where $\mu_{i}'$ is the inverse mapping of $\mu_{i} : \sX_{i} \rightarrow \mu_{i}\left( \sX_{i} \right)$. Since $\hat{g}$ must be a polynomial function as shown, the statement is established.
\end{IEEEproof}

\begin{remark}\rm
Another proof of Lemma \ref{lma:polynomial} involving Fermat's little theorem can be found in \cite{Huang2012a}.
\end{remark}

The important message conveyed by Lemma \ref{lma:polynomial} says that any discrete function defined on a finite domain is essentially a \emph{restriction} \cite[Definition II.3]{Huang2012a} of some polynomial function. Therefore, we can restrict the consideration of Problem \ref{prb:2} to all polynomial functions. This polynomial approach\footnote{This polynomial approach is first proposed in \cite{Huang2012a, Huang2012b}.} offers a very good insight into the general problem. After all, the algebraic structure of a polynomial function is much more clear than an arbitrary mapping (function). Most importantly, a polynomial function can often be expressed in several formats. Some of them are very helpful in tackling Problem \ref{prb:2} \cite{Huang2012a, Huang2012b}.

\begin{lemma}\label{lma:pre_poly_func_presentation}
Let $\sX_{1}, \sX_{2}, \cdots, \sX_{s}$ and $\sY$ be some finite sets. For any discrete function $g : \prod_{t=1}^{s} \sX_{t} \rightarrow \sY$, there exist a finite ring (field) $\fR$, functions $k_{t} : \sX_{t} \rightarrow \fR$ and $h : \fR \rightarrow \sY$, such that
\begin{align}
g(x_{1}, x_{2}, \cdots, x_{s}) = h\left( \sum_{t=1}^{s} k_{t}(x_{t}) \right) \label{eq:0:poly_presentation}.
\end{align}
\end{lemma}

\begin{IEEEproof}
There are several proofs of this lemma. One is provided in appendix \ref{app:b}.
\end{IEEEproof}

We often name the polynomial function $\hat{g}$ in Lemma \ref{lma:polynomial} a \emph{polynomial presentation} of $g$. This paper mainly focuses on presentations of format \eqref{eq:0:poly_presentation}. Readers are kindly referred to \cite{Huang2012b} for other interested formats.
As a simple demonstration \cite{Huang2012a}, once can see that the function $\min\{ x, y \}$ defined on $\{ 0, 1 \} \times \{ 0, 1 \}$ (with order $0 < 1$) admits polynomial presentations $x y \in \IZ_{2}[2]$ and $x + y - (x + y)^{2}$ defined on $\{ 0, 1 \} \times \{ 0, 1 \} \subsetneq \IZ_{3}^{2}$. The second one is of format \eqref{eq:0:poly_presentation}.

\section{Stochastic Complement, Reduced Markov Chains and Supremus
  Typical Sequences}

Given a Markov chain $\sM = \left\{ X^{(n)}
\right\}_{-\infty}^{\infty}$ with state space $\sX$ and a non-empty
subset $A$ of $\sX$, let
\begin{align*}
T_{A, l} =
\begin{cases}
\inf\left\{ n > 0 | X^{(n)} \in A \right\}; & l = 1,\\
\inf\left\{ n > T_{A, l-1} | X^{(n)} \in A \right\}; & l > 1,\\
\sup\left\{ n < T_{A, l+1} | X^{(n)} \in A \right\}; & l < 1.
\end{cases}
\end{align*}
It is well-known that $\sM_{A} = \left\{ X^{(T_{A, l})} \right\}_{-\infty}^{\infty}$ is Markov by the strong Markov property \cite[Theorem 1.4.2]{Norris1998}. In particular, if $\sM$ is irreducible, so is $\sM_{A}$. To be more precise, if $\sM$ is irreducible, and write its invariant distribution and transition matrix as $\pi = [ p_{i} ]_{i \in \sX}$ and
\begin{align*}
\bP =
\begin{bmatrix}
\bP_{A, A} & \bP_{A, A^{c}}\\
\bP_{A^{c}, A} & \bP_{A^{c}, A^{c}}
\end{bmatrix},
\end{align*}
respectively, then
\begin{align*}
\bS_{A} = \bP_{A, A} + \bP_{A, A^{c}} \left( \mathbf{1} - \bP_{A^{c}, A^{c}} \right)^{-1} \bP_{A^{c}, A},
\end{align*}
is the transition matrix of $\sM_{A}$ \cite[Theorem 2.1 and Section 3]{Meyer1989}. $\pi_{A} = \left[ \dfrac{ p_{i} }{ \sum_{j \in A} p_{j} } \right]_{i \in A}$ is an invariant distribution of $\bS_{A}$, i.e. $\pi_{A} \bS_{A} = \pi_{A}$ \cite[Theorem 2.2]{Meyer1989}. Since $\sM_{A}$ inherits irreducibility from $\sM$ \cite[Theorem 2.3]{Meyer1989}, $\pi_{A}$ is unique. The matrix $\bS_{A}$ is termed the \emph{stochastic complement} of $\bP_{A, A}$ in $\bP$, while $\sM_{A}$ is named a \emph{reduced Markov chain} of $\sM$. It has state space $A$ obviously.
%

\begin{definition}[Supremus Typicality]\label{def:Sup_tp}\rm
  Following the notation defined above, given $\epsilon > 0$ and a
  sequence $\bx = \left[ x^{(1)}, x^{(2)}, \cdots, x^{(n)} \right] \in
  \sX^{n}$ of length $n$ ($\geq 2 \abs{\sX}$), let $\bx_{A}$ be the
  subsequence of $\bx$ formed by all those $x^{(l)}$'s that belong to
  $A$ in the original ordering. $\bx$ is said to be \emph{Supremus
    $\epsilon$-typical} with respect to $\bP$, if and only if
  $\bx_{A}$ is strongly Markov $\epsilon$-typical with respect to
  $\bS_{A}$ for any feasible non-empty subset $A$ of $\sX$.  The set
  of all Supremus $\epsilon$-typical sequences with respect to $\bP$
  in $\sX^{n}$ is denoted $\S_{\epsilon}(n, \bP)$ or $\S_{\epsilon}$
  for simplicity.
\end{definition}

\begin{proposition}[AEP of Supremus
  Typicality]\label{prop:aep_of_Sup_tp}
  Let $\sM = \left\{ X^{(n)} \right\}_{-\infty}^{\infty}$ be an
  irreducible Markov chain with state space $\sX$, and $\bP = [ p_{i,
    j} ]_{i, j \in \sX}$ and $\pi = \left[ p_{j} \right]_{j \in \sX}$
  be its transition matrix and invariant distribution, respectively.
  For any $\eta > 0$, there exist $\epsilon_{0} > 0$ and $N_{0} \in
  \IN^{+}$, such that, $\forall \; \epsilon_{0} > \epsilon > 0$,
  $\forall \; n > N_{0}$ and $\forall \; \bx = \left[ x^{(1)},
    x^{(2)}, \cdots, x^{(n)} \right] \in \S_{\epsilon}(n, \bP)$,
\begin{enumerate}
\item\label{itm:1:aep_of_sup_tp} $\exp_{2}\left[ - n \left( H(\bP | \pi) + \eta \right) \right] < \Prb{ \left[ X^{(1)}, X^{(2)}, \cdots, X^{(n)} \right] = \bx } < \exp_{2}\left[ - n \left( H(\bP | \pi) - \eta \right) \right]$;

\item\label{itm:2:aep_of_sup_tp} $\Prb{ \bX \notin \S_{\epsilon}(n, \bP) } < \eta$, where $\bX = \left[ X^{(1)}, X^{(2)}, \cdots, X^{(n)} \right]$; and

\item\label{itm:3:aep_of_sup_tp} $\abs{ \S_{\epsilon}(n, \bP) } < \exp_{2}\left[ n \left( H(\bP | \pi) + \eta \right) \right]$.
\end{enumerate}
\end{proposition}

\begin{IEEEproof}
  Note that $\T_{\epsilon}(n, \bP) \supseteq \S_{\epsilon}(n,
  \bP)$. Thus, \ref{itm:1:aep_of_sup_tp}) and
  \ref{itm:3:aep_of_sup_tp}) are inherited from the AEP of strongly
  Markov typicality. In addition, \ref{itm:2:aep_of_sup_tp}) can be
  proved without any difficulty since any reduced Markov chain of
  $\sM$ is irreducible and the number of reduced Markov chains of
  $\sM$ is, $2^{\abs{\sX}} - 1$, finite.
\end{IEEEproof}

\begin{remark}\rm
  Motivated by Definition \ref{def:Sup_tp}, Proposition
  \ref{prop:aep_of_Sup_tp} and two related typicality lemmas in
  Appendix \ref{app:d}, one can define the concept of \emph{Supremus
    type} resembling other classic types \cite{Csiszar1998},
  e.g. Markov type \cite{Davisson1981b}. We will consider this in our
  future work for inspecting  error exponents of the schemes
  introduced in this paper.
\end{remark}

The following are two typicality lemmas of Supremus typical sequences tailored for our discussions. They are the ring specials of the two given in Appendix \ref{app:d}, respectively.

\begin{lemma}\label{lma:diff_typical_set}
Let $\fR$ be a finite ring, $\sM = \left\{ X^{(n)} \right\}_{-\infty}^{\infty}$ be an irreducible Markov chain whose state space, transition matrix and invariant distribution are $\fR$, $\bP$ and $\pi = \left[ p_{j} \right]_{j \in \fR}$, respectively. For any $\eta > 0$, there exist $\epsilon_{0} > 0$ and $N_{0} \in \IN^{+}$, such that, $\forall \; \epsilon_{0} > \epsilon > 0$, $\forall \; n > N_{0}$, $\forall \; \bx \in \S_{\epsilon}(n, \bP)$ and $\forall \; \fI \leq_{l} \fR$,
\begin{align}
\abs{ S_{\epsilon}(\bx, \fI) } < & \exp_{2}\left\{ n \left[ \sum_{A \in \fR / \fI} \sum_{j \in A} p_{j} H(\bS_{A} | \pi_{A}) + \eta \right] \right\} \label{ineq:1:diff_typical_set}\\
= & \exp_{2}\left\{ n \left[ H (\bS_{\fR / \fI} | \pi) + \eta \right] \right\} \label{ineq:2:diff_typical_set}
\end{align}
where
\begin{align*}
S_{\epsilon}(\bx, \fI) = \left\{ \left. \by \in \S_{\epsilon}(n, \bP) \right| \by - \bx \in \fI^{n} \right\},
\end{align*}
$\bS_{A}$ is the stochastic complement of $\bP_{A, A}$ in $\bP$, $\pi_{A} = \left[ \dfrac{p_{i}}{\sum_{j \in A} p_{j}} \right]_{i \in A}$ is the invariant distribution of $\bS_{A}$ and
$$
\bS_{\fR / \fI} = \diag{ \left\{ \bS_{A} \right\}_{A \in \fR / \fI} }.
$$
\end{lemma}

\begin{IEEEproof}
Assume that $\bx = \left[ x^{(1)}, x^{(2)}, \cdots, x^{(n)} \right]$
and let $\bx_{A}$ be the subsequence of $\bx$ formed by all those
$x^{(l)}$'s that belong to $A \in \fR / \fI$ in the original ordering. For any $\by = \left[ y^{(1)}, y^{(2)}, \cdots, y^{(n)} \right] \in S_{\epsilon}(\bx, \fI)$, obviously $y^{(l)} \in A$ if and only if $x^{(l)} \in A$ for all $A \in \fR / \fI$ and $1 \leq l \leq n$. Let $\bx_{A} = \left[ x^{(n_{1})}, x^{(n_{2})}, x^{(n_{m_{A}})} \right]$ (note: $\sum_{A \in \fR / \fI} m_{A} = n$ and $\abs{ \dfrac{m_{A}}{n} - \sum_{j \in A} p_{j} } < \abs{A} \epsilon + \dfrac{1}{n}$). It is easily seen that $\by_{A} = \left[ y^{(n_{1})}, y^{(n_{2})}, y^{(n_{m_{A}})} \right] \in A^{m_{A}}$ is a strongly Markov $\epsilon$-typical sequence of length $m_{A}$ with respect to $\bS_{A}$, since $\by$ is Supremus $\epsilon$-typical. Additionally, by Proposition \ref{prop:aep_of_Mar_tp}, there exist $\epsilon_{A} > 0$ and positive integer $M_{A}$ such that the number of strongly Markov $\epsilon$-typical sequences of length $m_{A}$ is upper bounded by $\exp_{2}\left\{ m_{A} \left[ H(\bS_{A} | \pi_{A}) + \eta/2 \right] \right\}$ if $0 < \epsilon < \epsilon_{A}$ and $m_{A} > M_{A}$. Therefore, if $0 < \epsilon < \min_{A \in \fR / \fI} \epsilon_{A}$, $n > M = \max_{A \in \fR / \fI}\left\{ \dfrac{ 1 + M_{A} }{ \abs{\sum_{j \in A} p_{j} - \abs{A} \epsilon} } \right\}$ (this guarantees that $m_{A} > M_{A}$ for all $A \in \fR / \fI$), then
\begin{align*}
\abs{S_{\epsilon}(\bx, \fI)} \leq & \exp_{2}\left\{ \sum_{A \in \fR / \fI} m_{A} \left[ H(\bS_{A} | \pi_{A}) + \eta/2 \right] \right\}\\
= & \exp_{2}\left\{ n \left[ \sum_{A \in \fR / \fI} \dfrac{m_{A}}{n} H(\bS_{A} | \pi_{A}) + \eta/2 \right] \right\}.
\end{align*}
Furthermore, choose $0 < \epsilon_{0} \leq \min_{A \in \fR / \fI} \epsilon_{A}$ and $N_{0} \geq M$ such that $\dfrac{m_{A}}{n} < \sum_{j \in A} p_{j} + \dfrac{\eta}{2 \sum_{A \in \fR / \fI} H(\bS_{A} | \pi_{A})}$ for all $0 < \epsilon < \epsilon_{0}$ and $n > N_{0}$ and $A \in \fR / \fI$, we have
\begin{align*}
\abs{S_{\epsilon}(\bx, \fI)} < & \exp_{2}\left\{ n \left[ \sum_{A \in \fR / \fI} \sum_{j \in A} p_{j} H(\bS_{A} | \pi_{A}) + \eta \right] \right\},
\end{align*}
\eqref{ineq:1:diff_typical_set} is established. Direct calculation yields \eqref{ineq:2:diff_typical_set}.
\end{IEEEproof}

\begin{lemma}\label{lma:diff_typical_set:ext}
In Lemma \ref{lma:diff_typical_set},
\begin{align}
\abs{ S_{\epsilon}(\bx, \fI) } < \exp_{2}\left\{ n \left[ H\left( \bP | \pi \right) - \lim_{m \rightarrow \infty} \dfrac{1}{m} H\left( Y_{\fR / \fI}^{(m)}, Y_{\fR / \fI}^{(m-1)}, \cdots, Y_{\fR / \fI}^{(1)} \right) + \eta \right] \right\}, \label{ineq:1:diff_typical_set:ext}
\end{align}
where $Y_{\fR / \fI}^{(m)} = X^{(m)} + \fI$ is a random variable with sample space $\fR / \fI$.
\end{lemma}

\begin{IEEEproof}
Assume that $\bx = \left[ x^{(1)}, x^{(2)}, \cdots, x^{(n)} \right]$ and let $$\overline{\by} = \left[ x^{(1)} + \fI, x^{(2)} + \fI, \cdots, x^{(n)} + \fI \right].$$
For any $\by = \left[ y^{(1)}, y^{(2)}, \cdots, y^{(n)} \right] \in S_{\epsilon}(\bx, \fI)$, obviously $y^{(l)} \in A$ if and only if $x^{(l)} \in A$ for all $A \in \fR / \fI$ and $1 \leq l \leq n$. Moreover, $$\overline{\by} = \left[ y^{(1)} + \fI, y^{(2)} + \fI, \cdots, y^{(n)} + \fI \right].$$
$\by$ is \emph{jointly typical} \cite{Cover1975} with $\overline{\by}$ with respect to the process
\begin{align*}
\cdots,
\left(\begin{matrix}
X^{(1)}\\
Y_{\fR / \fI}^{(1)}
\end{matrix}\right),
\left(\begin{matrix}
X^{(2)}\\
Y_{\fR / \fI}^{(2)}
\end{matrix}\right),
\cdots,
\left(\begin{matrix}
X^{(n)}\\
Y_{\fR / \fI}^{(n)}
\end{matrix}\right),
\cdots
\end{align*}
Therefore, there exist $\epsilon_{0} > 0$ and $N_{0} \in \IN^{+}$, such that, $\forall \; \epsilon_{0} > \epsilon > 0$ and $\forall \; n > N_{0}$,
\begin{align*}
\abs{ S_{\epsilon}(\bx, \fI) } < & \exp_{2}\bigg\{ n \bigg[ \lim_{m \rightarrow \infty} \dfrac{1}{m} H\left( X^{(m)}, X^{(m-1)}, \cdots, X^{(1)} \right)\\
 & - \lim_{m \rightarrow \infty} \dfrac{1}{m} H\left( Y_{\fR / \fI}^{(m)}, Y_{\fR / \fI}^{(m-1)}, \cdots, Y_{\fR / \fI}^{(1)} \right) + \eta \bigg] \bigg\}\\
= & \exp_{2}\bigg\{ n \bigg[ H\left( \bP | \pi \right) - \lim_{m \rightarrow \infty} \dfrac{1}{m} H\left( Y_{\fR / \fI}^{(m)}, Y_{\fR / \fI}^{(m-1)}, \cdots, Y_{\fR / \fI}^{(1)} \right) + \eta \bigg] \bigg\},
\end{align*}
where the equality follows from the fact that $\lim_{m \rightarrow \infty} \dfrac{1}{m} H\left( X^{(m)}, X^{(m-1)}, \cdots, X^{(1)} \right) = H\left( \bP \left| \pi \right. \right)$ since $\sM$ is irreducible Markov.
\end{IEEEproof}

\begin{remark}\rm
In Lemma \ref{lma:diff_typical_set:ext}, if $\bP = c_{1} \bU + ( 1 - c_{1} ) \mathbf{1}$ with all rows of $\bU$ being identical and $0 \leq c_{1} \leq 1$, then $\sM' = \left\{ Y_{\fR / \fI}^{(n)} \right\}_{-\infty}^{\infty}$ is Markovian by Lemma \ref{lma:Markovian_func}. As a conclusion,
\begin{align}
\abs{ S_{\epsilon}(\bx, \fI) } < & \exp_{2}\left\{ n \left[ H\left( \bP | \pi \right) - \lim_{m \rightarrow \infty} H\left( Y_{\fR / \fI}^{(m)} \left| Y_{\fR / \fI}^{(m-1)} \right. \right) + \eta \right] \right\} \label{ineq:2:diff_typical_set:ext}\\
= & \exp_{2}\left\{ n \left[ H\left( \bP | \pi \right) - H\left( \bP' | \pi' \right) + \eta \right] \right\}, \nonumber
\end{align}
where $\bP'$ and $\pi'$ are the transition matrix and the invariant distribution of $\sM'$ that can be easily calculated from $\bP$.
However, in general $\sM'$ is ergodic, but not Markovian. Its \emph{entropy rate} is difficult to obtain.
\end{remark}

\begin{remark}\rm
If $\fR$ in Lemma \ref{lma:diff_typical_set} is a field, then both \eqref{ineq:2:diff_typical_set} and \eqref{ineq:1:diff_typical_set:ext} are equivalent to
$$
\abs{ S_{\epsilon}(\bx, \fI) } < \exp_{2}\left[ n \left( H\left( \bP | \pi \right) + \eta \right) \right].
$$
Or, if $\sM$ in Lemma \ref{lma:diff_typical_set} is i.i.d., then both \eqref{ineq:2:diff_typical_set} and \eqref{ineq:1:diff_typical_set:ext} are equivalent to
$$
\abs{ S_{\epsilon}(\bx, \fI) } < \exp_{2}\left[ n \left( H\left( X^{(1)} \right) - H\left( Y_{\fR / \fI}^{(1)} \right) + \eta \right) \right],
$$
which is a special case of the \emph{generalized conditional typicality lemma} \cite[Lemma III.5]{Huang2012d}.
However, it is hard to determine which bound of these two is tighter in general.
Nevertheless, \eqref{ineq:2:diff_typical_set} is seemingly easier to analyze, while \eqref{ineq:1:diff_typical_set:ext} is more complicated for associating with the entropy rate of the ergodic process $\left\{ Y_{\fR / \fI}^{(n)} \right\}_{-\infty}^{\infty}.$
\end{remark}

\begin{remark}\rm
Lemma \ref{lma:diff_typical_set} and Lemma \ref{lma:diff_typical_set:ext} can be easily generalized to corresponding versions regarding other algebraic structures, e.g. group, rng\footnote{A ring without multiplicative identity.}, vector space, module, algebra and etc.
\end{remark}

\section{Achievability Theorem of Linear Coding for One Markov Source}

Equipped with the foundation laid down by Lemma \ref{lma:diff_typical_set} and Lemma \ref{lma:diff_typical_set:ext}, we resume our discussion to Problem \ref{prb:2}. For the time being, this section only considers a special scenario, namely $s = 1$, $g$ is an identity function and $\sM = \left\{ X_{1}^{(n)} \right\}_{-\infty}^{\infty} = \left\{ Y^{(n)} \right\}_{-\infty}^{\infty}$ is irreducible Markov in Problem \ref{prb:2}. It is known from \cite{Cover1975} that the achievable coding rate region for such a scenario is $\left\{ R \in \IR | R > H(\bP | \pi) \right\}$ where $\bP$ and $\pi$ are the transition matrix and invariant distribution of $\sM$, respectively. Unfortunately, the structures of the encoders used in \cite{Cover1975} are unclear which limits their application (to Problem \ref{prb:2}) as we will see in later sections. This motivates our study of encoders with explicit algebraic structures. We will examine the achievability of linear encoder over a finite ring for this special scenario of Problem \ref{prb:2}. The significance of this to other more general settings of Problem \ref{prb:2}, where $s$ and $g$ are both arbitrary, will be seen in Section \ref{sec:computing_Mar_func}.

\begin{theorem}\label{thm:1:achievability}
Assume that $s = 1$, $\sX_{1} = \sY$ is some finite ring $\fR$ and $g$ is an identity function in Problem \ref{prb:2}, and additionally $\left\{ X_{1}^{(n)} \right\}_{-\infty}^{\infty} = \left\{ Y^{(n)} \right\}_{-\infty}^{\infty}$ is irreducible Markov with transition matrix $\bP$ and invariant distribution $\pi$.
We have that
\begin{align}
R > \max_{0 \neq \fI \leq_{l} \fR} \dfrac{\log \abs{\fR}}{\log \abs{\fI}} \min\left\{ H(\bS_{\fR / \fI} | \pi), H\left( \bP | \pi \right) - \lim_{m \rightarrow \infty} \dfrac{1}{m} H\left( Y_{\fR / \fI}^{(m)}, Y_{\fR / \fI}^{(m-1)}, \cdots, Y_{\fR / \fI}^{(1)} \right) \right\}, \label{eq:achievability}
\end{align}
where
$$
\bS_{\fR / \fI} = \diag{ \left\{ \bS_{A} \right\}_{A \in \fR / \fI} }
$$
with $\bS_{A}$ being the stochastic complement of $\bP_{A, A}$ in
$\bP$ and $Y_{\fR / \fI}^{(i)} = X_{1}^{(i)} + \fI$, is achievable
with linear coding over $\fR$.  To be more precise, for any $\epsilon
> 0$, there is an $N_{0} \in \IN^{+}$ such that there exist a linear
encoder $\phi : \fR^{n} \rightarrow \fR^{k}$ and a decoder $\psi :
\fR^{k} \rightarrow \fR^{n}$ for all $n > N_{0}$ with
\begin{align*}
\Prb{ \psi\left( \phi\left( Y^{n} \right) \right) \neq Y^{n} } < \epsilon,
\end{align*}
provided that
\begin{align*}
k > \max_{0 \neq \fI \leq_{l} \fR} \dfrac{n}{\log \abs{\fI}} \min\left\{ H(\bS_{\fR / \fI} | \pi), H\left( \bP | \pi \right) - \lim_{m \rightarrow \infty} \dfrac{1}{m} H\left( Y_{\fR / \fI}^{(m)}, Y_{\fR / \fI}^{(m-1)}, \cdots, Y_{\fR / \fI}^{(1)} \right) \right\}.
\end{align*}
\end{theorem}

Generally speaking, $\sX$ or $\sY$ is not necessarily associated with any algebraic structure. In order to apply the linear encoder, we usually assume that $\sY$ in Problem \ref{prb:2} is mapped into a finite ring $\fR$ of order at least $\abs{\sY}$ by some injection $\Phi : \sY \rightarrow \fR$ and denote the set of all possible injections by $\I(\sY, \fR)$.

\begin{theorem}\label{thm:2:achievability}
Assume that $s = 1$, $g$ is an identity function and $\left\{ X_{1}^{(n)} \right\}_{-\infty}^{\infty} = \left\{ Y^{(n)} \right\}_{-\infty}^{\infty}$ is irreducible Markov with transition matrix $\bP$ and invariant distribution $\pi$ in Problem \ref{prb:2}. For a finite ring $\fR$ of order at least $\abs{\sY}$ and $\forall \; \Phi \in \I(\sY, \fR)$, let
\begin{align*}
r_{\Phi} = \max_{0 \neq \fI \leq_{l} \fR} \dfrac{\log \abs{\fR}}{\log \abs{\fI}} \min\left\{ H(\bS_{\Phi, \fI} | \pi), H\left( \bP | \pi \right) - \lim_{m \rightarrow \infty} \dfrac{1}{m} H\left( Y_{\fR / \fI}^{(m)}, Y_{\fR / \fI}^{(m-1)}, \cdots, Y_{\fR / \fI}^{(1)} \right) \right\},
\end{align*}
where
$$
\bS_{\Phi, \fI} = \diag{ \left\{ \bS_{\Phi^{-1}(A)} \right\}_{A \in \fR / \fI} }
$$
with $\bS_{\Phi^{-1}(A)}$ being the stochastic complement of $\bP_{\Phi^{-1}(A), \Phi^{-1}(A)}$ in $\bP$ and $Y_{\fR / \fI}^{(m)} = \Phi\left( X_{1}^{(m)} \right) + \fI$,
and define
\begin{align*}
\R_{\Phi} = \left\{ R \in \IR | R > r_{\Phi} \right\}.
\end{align*}
We have that
\begin{align}
\bigcup_{\Phi \in \I(\sY, \fR)} \R_{\Phi} \label{eq:conv_hull:achievability}
\end{align}
is achievable with linear coding over $\fR$.
\end{theorem}

\begin{IEEEproof}
The result follows immediately from Theorem \ref{thm:1:achievability}.
\end{IEEEproof}

\begin{remark}\rm
In Theorem \ref{thm:2:achievability}, assume that $\sY$ is some finite ring itself, and let $\tau$ be the identity mapping in $\I(\sY, \sY)$. It could happen that $\R_{\tau} \subsetneq \R_{\Phi}$ for some $\Phi \in \I(\sY, \sY)$. This implies that region given by \eqref{eq:achievability} can be strictly smaller than \eqref{eq:conv_hull:achievability}. Therefore, a ``reordering'' of elements in the ring $\sY$ is required when seeking for better linear encoders.
\end{remark}

\begin{remark}\rm
By Lemma \ref{lma:Markovian_func}, if, in Theorem \ref{thm:1:achievability}, $\bP = c_{1} \bU + ( 1 - c_{1} ) \mathbf{1}$ with all rows of $\bU$ being identical and $0 \leq c_{1} \leq 1$, then
\begin{align*}
R > \max_{0 \neq \fI \leq_{l} \fR} \dfrac{\log \abs{\fR}}{\log \abs{\fI}} \min\left\{ H(\bS_{\fR / \fI} | \pi), H\left( \bP | \pi \right) - \lim_{m \rightarrow \infty} H\left( Y_{\fR / \fI}^{(m)} \left| Y_{\fR / \fI}^{(m-1)} \right. \right) \right\}
\end{align*}
is achievable with linear coding over $\fR$. Similarly, if $\bP = c_{1} \bU + ( 1 - c_{1} ) \mathbf{1}$ in Theorem \ref{thm:2:achievability}, then, for all $\Phi \in \I(\sY, \fR)$,
\begin{align*}
\R_{\Phi} = \left\{ R \in \IR \Big| R > \max_{0 \neq \fI \leq_{l} \fR} \dfrac{\log \abs{\fR}}{\log \abs{\fI}} \min\left\{ H(\bS_{\Phi, \fI} | \pi), H\left( \bP | \pi \right) - \lim_{m \rightarrow \infty} H\left( Y_{\fR / \fI}^{(m)} \left| Y_{\fR / \fI}^{(m-1)} \right. \right) \right\} \right\}.
\end{align*}
\end{remark}

\begin{IEEEproof}[Proof of Theorem \ref{thm:1:achievability}]
Let
\begin{align*}
R_{0} = \max_{0 \neq \fI \leq_{l} \fR} \dfrac{\log \abs{\fR}}{\log \abs{\fI}} \min\left\{ H(\bS_{\fR / \fI} | \pi), H\left( \bP | \pi \right) - \lim_{m \rightarrow \infty} \dfrac{1}{m} H\left( Y_{\fR / \fI}^{(m)}, Y_{\fR / \fI}^{(m-1)}, \cdots, Y_{\fR / \fI}^{(1)} \right) \right\}
\end{align*}
and, for any $R > R_{0}$ and $n \in \IN^{+}$, let $k = \left\lfloor \dfrac{n R}{ \log \abs{\fR} } \right\rfloor$. Obviously, there always exists $N_{0}' \in \IN^{+}$ such that, for any $0 \neq \fI \leq_{l} \fR$ and $\dfrac{\log \abs{\fI}}{\log \abs{\fR}} \dfrac{R - R_{0}}{2} > \eta > 0$,
\begin{align}
\min\left\{ H(\bS_{\fR / \fI} | \pi), H\left( \bP | \pi \right) - \lim_{m \rightarrow \infty} \dfrac{1}{m} H\left( Y_{\fR / \fI}^{(m)}, Y_{\fR / \fI}^{(m-1)}, \cdots, Y_{\fR / \fI}^{(1)} \right) \right\} + \eta - \dfrac{k}{n} \log \abs{\fI} < & - \eta/2 \label{eq:0:achievability}
\end{align}
if $n > N_{0}'$. The following proves that $R$ is achievable with linear coding over $\fR$.

\begin{enumerate}
\item\label{itm:1:achievability} Encoding:

Choose some $n \in \IN^{+}$ and generate a $k \times n$ matrix $\bA$ over $\fR$ uniformly at  random (independently choose each entry of $\bA$ from $\fR$ uniformly at random). Let the encoder be the linear mapping
\begin{align*}
\phi : \bx \mapsto \bA \bx, \forall \; \bx \in \fR^{n}.
\end{align*}
We note that the coding rate is $\dfrac{1}{n} \log \abs{\phi(\fR^{n})} \leq \dfrac{1}{n} \log \abs{\fR^{k}} = \dfrac{\log \abs{\fR}}{n} \left\lfloor \dfrac{n R}{ \log \abs{\fR} } \right\rfloor \leq R$.

\item Decoding:

Choose an $\epsilon > 0$. Assume that $\bz \in \fR^{k}$ is the observation, the decoder claims that $\bx \in \fR^{n}$ is the original data sequence encoded, if and only if
\begin{enumerate}
\item
$\bx \in \S_{\epsilon}(n, \bP)$; and

\item
$\forall \; \bx' \in \S_{\epsilon}(n, \bP)$, if $\bx' \neq \bx$, then $\phi(\bx') \neq \bz$. In other words, the decoder $\psi$ maps $\bz$ to $\bx$.

\end{enumerate}

\item Error:

Assume that $\bX \in \fR^{n}$ is the original data sequence generated. An error occurs if and only if
\begin{enumerate}
\item[$E_{1}$]
$\bX \notin \S_{\epsilon}(n, \bP)$; or

\item[$E_{2}$]
There exists $\bx' \in \S_{\epsilon}(n, \bP)$ such that $\phi(\bx') = \phi(\bX)$.

\end{enumerate}

\item\label{itm:4:achievability} Error Probability:

We claim that there exist $N_{0} \in \IN^{+}$ and $\epsilon_{0} > 0$, if $n > N_{0}$ and $\epsilon_{0} > \epsilon > 0$, then $\Prb{ \psi( \phi(\bX) ) \neq \bX } = \Prb{ E_{1} \cup E_{2} } < \eta$.
First of all, by the AEP of Supremus typicality (Proposition \ref{prop:aep_of_Sup_tp}), there exist $N_{0}'' \in \IN^{+}$ and $\epsilon_{0}'' > 0$ such that $\Prb{ E_{1} } < \eta/2$ if $n > N_{0}''$ and $\epsilon_{0}'' > \epsilon > 0$.
Secondly, let $E_{1}^{c}$ be the complement of $E_{1}$. We have
\begin{align}
 & \cPrb{ E_{2} }{ E_{1}^{c} } \nonumber\\
= & \sum_{\bx' \in \S_{\epsilon} \setminus \{\bX\}} \cPrb{ \phi(\bx') = \phi(\bX) }{ E_{1}^{c} } \nonumber\\
\leq & \sum_{0 \neq \fI \leq_{l} \fR} \; \sum_{\bx' \in S_{\epsilon}(\bX, \fI) \setminus \{\bX\}} \cPrb{ \phi(\bx') = \phi(\bX) }{ E_{1}^{c} } \label{eq:1:achievability}\\
< & \sum_{0 \neq \fI \leq_{l} \fR} \exp_{2}\left[ n ( r_{\fR / \fI} + \eta ) \right] \abs{\fI}^{-k} \label{eq:2:achievability}\\
< & \left( 2^{\abs{\fR}} - 2 \right) \max_{0 \neq \fI \leq_{l} \fR} \exp_{2}\left[ n \left( r_{\fR / \fI} + \eta - \dfrac{k}{n} \log \abs{\fI} \right) \right] \label{eq:3:achievability}\\
< & \left( 2^{\abs{\fR}} - 2 \right) \exp_{2}( - n \eta/ 2), \label{eq:4:achievability}
\end{align}
where $r_{\fR / \fI} = \min\left\{ H(\bS_{\fR / \fI} | \pi), H\left( \bP | \pi \right) - \lim_{m \rightarrow \infty} \dfrac{1}{m} H\left( Y_{\fR / \fI}^{(m)}, Y_{\fR / \fI}^{(m-1)}, \cdots, Y_{\fR / \fI}^{(1)} \right) \right\}$,
\begin{enumerate}
\item[\eqref{eq:1:achievability}]
follows from the fact that $\S_{\epsilon}(n, \bP) = \bigcup_{0 \neq \fI \leq_{l} \fR} S_{\epsilon}(\bX, \fI)$;

\item[\eqref{eq:2:achievability}]
is from the typicality lemmas, Lemma \ref{lma:diff_typical_set} and Lemma \ref{lma:diff_typical_set:ext}, and \cite[Lemma III.3]{Huang2012d}, and it is required that $\epsilon$ is smaller than some $\epsilon_{0}''' > 0$ and $n$ is larger than some $N_{0}''' \in \IN^{+}$;

\item[\eqref{eq:3:achievability}]
is due to the fact that the number of non-trivial left ideals of $\fR$ is bounded by $2^{\abs{\fR}} - 2$;

\item[\eqref{eq:4:achievability}] is from \eqref{eq:0:achievability}, and it is required that $n > N_{0}'$.

\end{enumerate}
Let $N_{0} = \max\left\{ N_{0}', N_{0}'', N_{0}''', \left\lceil \dfrac{2}{\eta} \log \left[ \dfrac{2}{\eta} \left( 2^{\abs{\fR}} - 2 \right) \right] \right\rceil \right\}$ and $\epsilon_{0} = \min\{ \epsilon_{0}'', \epsilon_{0}''' \}$. We have that
\begin{align*}
\cPrb{ E_{2} }{ E_{1}^{c} } < \eta/2 \mbox{ and } \Prb{ E_{1}^{c} } < \eta/2
\end{align*}
if $n > N_{0}$ and $\epsilon_{0} > \epsilon > 0$. Hence,
$\Prb{E_{1} \cup E_{2}} = \cPrb{ E_{2} }{ E_{1}^{c} } + \Prb{
  E_{1}^{c} } < \eta$.

\end{enumerate}

By \ref{itm:1:achievability}) -- \ref{itm:4:achievability}), the theorem is established.
\end{IEEEproof}

\begin{remark}\rm
From the proof of Theorem \ref{thm:1:achievability} (\cite[Theorem III.1]{Huang2012d}), one can see that the generalization of the achievability theorem from linear coding technique over finite field to the one over finite ring builds on the generalization of the typicality lemma of Markov sources (the conditional typicality lemma of correlated i.i.d. sources \cite[Theorem 15.2.2]{Cover2006}) and the analysis of random linear mappings over finite rings \cite[Lemma III.3]{Huang2012d}.
\end{remark}

The following is an example to help interpreting the above theorems.
It is seen from this example that \eqref{eq:achievability}, as well as \eqref{eq:conv_hull:achievability}, coincides with \eqref{eq:cover_ergodic_sources} for $s = 1$.

\begin{example}\label{eg:achievability}\rm
Let $\sM$ be an irreducible Markov chain with state space $\IZ_{4} = \left\{ 0, 1, 2, 3 \right\}$. Its transition matrix $\bP = [ p_{i, j} ]_{i, j \in \IZ_{4}}$ is given as the follows.
\begin{center}
\begin{tabular}{| c | c | c | c | c |}
\hline
 & 0 & 1 & 2 & 3\\
\hline
0 & .8142 & .1773 & .0042 & .0042\\
\hline
1 & .0042 & .9873 & .0042 & .0042\\
\hline
2 & .0042 & .1773 & .8142 & .0042\\
\hline
3 & .0042 & .1773 & .0042 & .8142\\
\hline
\end{tabular}
\end{center}
By Theorem \ref{thm:1:achievability}, we have that $\R = \left\{ R \in \IR | R > \max\{ 0.1602, 0.1474 \} = H(\bP | \pi) \right\}$, where $\pi$ is the invariant distribution of $\sM$, is achievable with linear coding over $\IZ_{4}$. One can easily see that $\R$ is just the optimal region given by \eqref{eq:cover_ergodic_sources} for $s = 1$.
\end{example}

Although the achievable regions presented in the above theorems are comprehensive, they depict the optimal one in many situations, i.e. \eqref{eq:conv_hull:achievability} (or \eqref{eq:achievability}) is identical to \eqref{eq:cover_ergodic_sources} for $s = 1$. This has been demonstrated by Example \ref{eg:achievability} above, and more is shown in the following.

\begin{corollary}\label{crl:1:achievability}
In Theorem \ref{thm:1:achievability} (or Theorem \ref{thm:2:achievability}), if $\fR$ is a finite field, then
\begin{align*}
 & R > H(\bP | \pi)\\
(\mbox{or } & \R_{\Phi} = \left\{ R \in \IR \left| R > H(\bP | \pi) \right. \right\}, \forall \; \Phi \in \I(\sY, \fR),)
\end{align*}
is achievable with linear coding over $\fR$.
\end{corollary}

\begin{IEEEproof}
If $\fR$ is a finite field, then $\fR$ is the only non-trivial left ideal of itself. The statement follows, since $\bS_{\fR / \fR} = \bP$ ($\bS_{\Phi, \fR} = \bP$) and $H\left( Y_{\fR / \fR}^{(m)} \right) = 0$ for all feasible $m$.
\end{IEEEproof}
%

Corollary \ref{crl:1:achievability} says that linear coding over
finite fields is always optimal for the special case of Problem
\ref{prb:2} considered in this section. However, it is not yet
conclusively proved that linear coding over any non-field ring can be
equally optimal, other than shown in Example
\ref{eg:achievability}. Nevertheless, it has been proved that, in the
case of multiple i.i.d. correlated sources, there always exist
non-field rings over which linear coding is optimal
\cite{Huang2012e}. As a matter of fact, the single source scenario of
this assertion is included as a special case of Theorem
\ref{thm:2:achievability} (see Corollary \ref{crl:2:achievability}).

\begin{corollary}\label{crl:2:achievability}
  In Theorem \ref{thm:2:achievability}, if $\bP$ describes an
  i.i.d. process, i.e. the row vectors of $\bP$ are identical to $\pi = [
  p_{j} ]_{j \in \sY}$, then
\begin{align*}
\R_{\Phi} = \left\{ R \in \IR \left| R > \max_{0 \neq \fI \leq_{l} \fR} \dfrac{ \log \abs{\fR} }{ \log \abs{\fI} } \left[ H(\pi) - H(\pi_{\Phi, \fI}) \right] \right. \right\}, \forall \; \Phi \in \I(\sY, \fR),
\end{align*}
where $\pi_{\Phi, \fI} = \left[ \sum_{j \in \Phi^{-1}(A)} p_{j} \right]_{A \in \fR / \fI}$,
is achievable with linear coding over $\fR$.
In particular, if
\begin{enumerate}
\item\label{itm:1:achievability}
$\fR$ is a field; or

\item\label{itm:2:achievability}
$\fR$ contains one and only one proper non-trivial left ideal $\fI_{0}$ and $\abs{\fI_{0}} = \sqrt{ \abs{\fR} }$; or

\item\label{itm:3:achievability}
$\fR$ is a product ring of several rings satisfying condition \ref{itm:1:achievability}) or \ref{itm:2:achievability}),

\end{enumerate}
then $\bigcup_{\Phi \in \I(\sY, \fR)} \R_{\Phi}$ is the Slepian--Wolf region
$
\left\{ R \in \IR \left| R > H(\pi) \right. \right\}.
$
\end{corollary}

\begin{IEEEproof}
The first half of the statement follows from Theorem \ref{thm:2:achievability} by direct calculation. The second half is from \cite{Huang2012e}.
\end{IEEEproof}

\begin{remark}\rm
Concrete examples of the finite ring from Corollary
\ref{crl:2:achievability} includes, but are not limited to:
\begin{enumerate}
\item
$\IZ_{p}$, where $p$ is a prime, as a finite field;

\item
$\IZ_{p^{2}}$ and
$M_{L, p}  = \left\{ \left.
\begin{bmatrix}
x & 0\\
y & x
\end{bmatrix}
\right| x, y \in \IZ_{p} \right\},$
where $p$ is a prime;

\item
$M_{L, p_{1}} \times \IZ_{p_{2}}$, where $p_{1}$ and $p_{2}$ are primes.

\end{enumerate}
Since there always exists a prime $p$ with $p^{2} > \abs{\sY}$ in Theorem \ref{thm:2:achievability}, Corollary \ref{crl:2:achievability} guarantees that there always exist optimal linear encoders over some non-field ring, say $\IZ_{p^{2}}$ or $M_{L, p}$, if the source is i.i.d..
\end{remark}

As mentioned, Corollary \ref{crl:2:achievability} can be generalized
to the multiple sources scenario in a memoryless setting (see
\cite{Huang2012d, Huang2012e}). In exact terms, the Slepian--Wolf
region is always achieved with linear coding over some non-field
ring. Unfortunately, it is neither proved nor denied that
a corresponding existence conclusion for the (single or multivariate
\cite{Fung2002}) Markov source(s) scenario holds. Nevertheless,
Example \ref{eg:achievability}, Corollary \ref{crl:2:achievability}
and \cite{Huang2012e} do affirmatively support such an assertion to
their own extents\footnote{The authors conjecture that linear coding
  claims optimality in the discussed aspect of the problem. However,
  there may be a weakness in the technique used to obtain
  \eqref{eq:conv_hull:achievability}. This weakness prohibits full
  extraction of the capability of the linear encoder. Consequently, it
  could happen that \eqref{eq:conv_hull:achievability} is strictly
  smaller than \eqref{eq:cover_ergodic_sources} for $s = 1$ in some
  cases.}.

Even if it is unproved that linear coding over non-field ring is
optimal for the scenario of Problem \ref{prb:2} considered in this
section, it will be seen in later sections that linear coding over
non-field ring strictly outperforms its field counterpart in other
settings of the problem.

\section{Source Coding for Computing Markovian Functions}\label{sec:computing_Mar_func}

We are now ready to move on to a more general setting of Problem
\ref{prb:2}, where both $s$ and $g$ are arbitrary. We begin with
briefing the reader on our main idea with Example
\ref{eg:computing_Mar_func} in the following. This example shows that
the achievable coding rate region for computing a linear function $g$
of $s$ variables is likely to be strictly larger than $\R_{s}$ in the
setting of sources with
memory.

\begin{example}\label{eg:computing_Mar_func}\rm
Consider three sources $S_{1}$, $S_{2}$ and $S_{3}$ generating random data $X_{1}^{(i)}$, $X_{2}^{(i)}$ and $X_{3}^{(i)}$ (at time $i \in \IN^{+}$) whose sample spaces are all $\sX_{1} = \sX_{2} = \sX_{3} = \{ 0, 1 \} \subsetneq \IZ_{4}$, respectively. Let $g : \sX_{1} \times \sX_{2} \times \sX_{3} \rightarrow \IZ_{4}$ be defined as
\begin{align}
g : (x_{1}, x_{2}, x_{3}) \mapsto x_{1} + 2 x_{2} + 3 x_{3}, \label{eq:1:example}
\end{align}
and assume that $\left\{ X^{(n)} \right\}_{-\infty}^{\infty}$, where $X^{(i)} = \left( X_{1}^{(i)}, X_{2}^{(i)}, X_{3}^{(i)} \right)$, forms a Markov chain with transition matrix
\begin{align*}
\begin{tabular}{| c | c | c | c | c | c | c | c | c |}
\hline
 & (0, 0, 0) & (0, 0, 1) & (0, 1, 0) & (0, 1, 1) & (1, 0, 0) & (1, 0, 1) & (1, 1, 0) & (1, 1, 1)\\
 \hline
(0, 0, 0) & .1397 & .4060 & .0097 & .0097 & .0097 & .0097 & .4060 & .0097\\
\hline
(0, 0, 1) & .0097 & .5360 & .0097 & .0097 & .0097 & .0097 & .4060 & .0097\\
\hline
(0, 1, 0) & .0097 & .4060 & .1397 & .0097 & .0097 & .0097 & .4060 & .0097\\
\hline
(0, 1, 1) & .0097 & .4060 & .0097 & .1397 & .0097 & .0097 & .4060 & .0097\\
\hline
(1, 0, 0) & .0097 & .4060 & .0097 & .0097 & .1397 & .0097 & .4060 & .0097\\
\hline
(1, 0, 1) & .0097 & .4060 & .0097 & .0097 & .0097 & .1397 & .4060 & .0097\\
\hline
(1, 1, 0) & .0097 & .4060 & .0097 & .0097 & .0097 & .0097 & .5360 & .0097\\
\hline
(1, 1, 1) & .0097 & .4060 & .0097 & .0097 & .0097 & .0097 & .4060 & .1397\\
\hline
\end{tabular}
\end{align*}

In order to recover $g$ at the decoder, one solution is to apply Cover's method \cite{Cover1975} to first decode the original data and then compute $g$. This results in an achievable region
\begin{align*}
\R_{3} = \bigg\{ [ R_{1}, R_{2}, R_{3} ] \in \IR^{3} \bigg| \sum_{t \in T} R_{t} > & \lim_{m \rightarrow \infty} \Big[ H\left( X_{1}^{(m)}, X_{2}^{(m)}, X_{3}^{(m)} \left| X_{1}^{(m-1)}, X_{2}^{(m-1)}, X_{3}^{(m-1)} \right. \right)\\
 & - H\left( X_{T^{c}}^{(m)} \left| X_{T^{c}}^{(m-1)} \right. \right) \Big], \emptyset \neq T \subseteq \left\{ 1, 2, 3 \right\} \bigg\}.
\end{align*}

However, $\R_{3}$ is not optimal, i.e. coding rates beyond this region can be achieved. Observe that $\left\{  Y^{(n)} \right\}_{-\infty}^{\infty}$, where $Y^{(i)} = g \left( X^{(i)} \right)$, is an irreducible Markovian with transition matrix
\begin{align}
\begin{tabular}{| c | c | c | c | c |}
\hline
 & 0 & 3 & 2 & 1\\
\hline
0 & .1493 & .8120 & .0193 & .0193\\
\hline
3 & .0193 & .9420 & .0193 & .0193\\
\hline
2 & .0193 & .8120 & .1493 & .0193\\
\hline
1 & .0193 & .8120 & .0193 & .1493\\
\hline
\end{tabular}\label{eq:1:eg_computing_Mar_func}
\end{align}
By Theorem \ref{thm:1:achievability}, for any $\epsilon > 0$, there
is an $N_{0} \in \IN^{+}$, such that for all $n > N_{0}$ there exist a
linear encoder $\phi : \IZ_{4}^{n} \rightarrow \IZ_{4}^{k}$ and a
decoder $\psi : \IZ_{4}^{k} \rightarrow \IZ_{4}^{n}$, such that $\Prb{
  \psi\left( \phi\left( Y^{n} \right) \right) \neq Y^{n} } <
\epsilon$, where $Y^{n} = \left[ Y^{(1)}, Y^{(2)}, \cdots, Y^{(n)}
\right]$, as long as
\begin{align*}
k > \dfrac{n}{ 2 } \times \max\left\{ 0.3664, 0.3226 \right\} = 0.1832 n.
\end{align*}
Further notice that
\begin{align*}
\phi\left( Y^{n} \right) = \vec{g}\left( Z_{1}^{k}, Z_{2}^{k}, Z_{3}^{k} \right),
\end{align*}
where
$Z_{t}^{k} = \phi\left( X_{t}^{n} \right)$ ($t = 1, 2, 3$) and
$\vec{g}\left( Z_{1}^{k}, Z_{2}^{k}, Z_{3}^{k} \right) =
\begin{bmatrix}
g\left( Z_{1}^{(1)}, Z_{2}^{(1)}, Z_{3}^{(1)} \right)\\
g\left( Z_{1}^{(2)}, Z_{2}^{(2)}, Z_{3}^{(2)} \right)\\
\vdots\\
g\left( Z_{1}^{(k)}, Z_{2}^{(k)}, Z_{3}^{(k)} \right)
\end{bmatrix},
$
since $g$ is also linear.
Thus, another approach\footnote{The idea of this approach is first introduced by K\"{o}rner and Marton \cite{Korner1979} for computing the modulo-two sum of two correlated i.i.d. sources. This is then generalized to the case of arbitrary discrete function based on the observation that any discrete function of finite domain is a restriction of some polynomial function over some finite field \cite{Huang2012a, Huang2012b}. The supports of these approaches are linear coding techniques over finite fields from Elias \cite{Elias1955} (binary field) and Csisz\'{a}r \cite{Csiszar1982} (arbitrary finite field). However, \cite{Huang2012d} points out that treating an arbitrary discrete function as a polynomial function over some finite ring (instead over field) can lead to strictly better performance. This (encoding polynomial functions over finite rings) requires establishing the achievability theorems, \cite[Theorem III.1]{Huang2012d} and Theorem \ref{thm:1:achievability}, of linear coding techniques over rings.}
is to use $\phi$ as encoder for each source. Upon observing
$Z_{1}^{k}$, $Z_{2}^{k}$ and $Z_{3}^{k}$, the decoder claims that
$\psi\left( \vec{g}\left( Z_{1}^{k}, Z_{2}^{k}, Z_{3}^{k} \right)
\right)$ is the desired data $\vec{g}\left( X_{1}^{n}, X_{2}^{n},
  X_{3}^{n} \right)$. Obviously 
\begin{align*}
 & \Prb{ \psi\left( \vec{g}\left[ \phi\left( X_{1}^{n} \right), \phi\left( X_{2}^{n} \right), \phi\left( X_{3}^{n} \right) \right] \right) \neq Y^{n} }\\
= & \Prb{ \psi\left( \phi\left( Y^{n} \right) \right) \neq Y^{n} } < \epsilon,
\end{align*}
as long as $k > 0.1832 n$.
As a consequence, the region
\begin{align}
\R_{\IZ_{4}} = \left\{ [ r, r, r ] \in \IR^{3} \left| r > \dfrac{2 k}{n} = 0.4422 \right. \right\} \label{eq:2:eg_computing_Mar_func}
\end{align}
is achieved. Since
$$
0.4422 + 0.4422 + 0.4422 < \lim_{m \rightarrow \infty} H\left( X_{1}^{(m)}, X_{2}^{(m)}, X_{3}^{(m)} \left| X_{1}^{(m-1)}, X_{2}^{(m-1)}, X_{3}^{(m-1)} \right. \right) = 1.4236,
$$
we have that $\R_{\IZ_{4}}$ is larger than $\R_{3}$. In conclusion, $\R_{3}$ is suboptimal for computing $g$.
\end{example}

Compared to the one stated in Example \ref{eg:computing_Mar_func}, the
native Problem \ref{prb:2} is too arbitrary in the sense that even the
stochastic property of the sources is unspecified. In order to obtain
meaningful conclusions, we will further assume that either condition
(c0) or condition (c1) holds. It is easy to see that Example
\ref{eg:computing_Mar_func} falls in the category of (c0) which is in
fact a special subclass of (c1). One practical interpretation of the
mechanism (c0) illustrates is as the following:
\begin{enumerate}
\item[] The datum generated at time $n + 1$ ($n \in \IN^{+}$) by each
  source inclines to be the same as the one generated at time
  $n$. However, due to some ``interference'' casted by the system, the
  generated data can vary based on a distribution $[ u_{x} ]_{x \in
    \sX}$ (a unitary vector). The weights of the two impacts are
  quantified by $1 - c_{1}$ and $c_{1}$, respectively.
\end{enumerate}
As a special case of (c0), if $c_{1} = 1$, then the generated data
sequence forms a correlated i.i.d. process. On the other hand, the
scene described by (c1) is much broader as mentioned. For instance,
$g$ can be a sum of two sources with non-ergodic stochastic
behavior, while the sum itself is Markovian. A very interesting
realization of such a phenomenon is given later in Example
\ref{eg:computing_Mar_func:ext}.

In the rest of this section, we will address (c1) first. The
conclusion for (c0) will then follow very naturally after the
connection between these two conditions is further detailed.

\begin{theorem}\label{thm:computing_c1_func}
In Problem \ref{prb:2}, assume that $g$ satisfies (c1), and let $\bP$ and $\pi$ be the transition matrix and invariant distribution of $\left\{ Z^{(n)} = \sum_{t \in \S} k_{t}\left( X_{t}^{(n)} \right) \right\}_{-\infty}^{\infty}$, respectively.
We have
\begin{align*}
\R = \left\{ [R, R, \cdots, R] \in \IR^{s} | R > R_{0} \right\} \subseteq \R[g],
\end{align*}
where
\begin{align*}
R_{0} = \max_{0 \neq \fI \leq_{l} \fR} \dfrac{\log \abs{\fR}}{\log \abs{\fI}} \min\left\{ H(\bS_{\fR / \fI} | \pi), H\left( \bP | \pi \right) - \lim_{m \rightarrow \infty} \dfrac{1}{m} H\left( Y_{\fR / \fI}^{(m)}, Y_{\fR / \fI}^{(m-1)}, \cdots, Y_{\fR / \fI}^{(1)} \right) \right\},
\end{align*}
$\bS_{\fR / \fI} = \diag{ \left\{ \bS_{A} \right\}_{A \in \fR / \fI} }$ with $\bS_{A}$ being the stochastic complement of $\bP_{A, A}$ in $\bP$ and $Y_{\fR / \fI}^{(m)} = Z^{(m)} + \fI$. Moreover, if $\fR$ is a field, then
\begin{align}
\R = \left\{ [R, R, \cdots, R] \in \IR^{s} \left| R > H(\bP | \pi) \right. \right\}. \label{eq:computing_c1_func}
\end{align}
\end{theorem}

\begin{IEEEproof}
By Theorem \ref{thm:1:achievability}, for any $\epsilon > 0$, there exists an $N_{0} \in \IN^{+}$ and for all $n > N_{0}$, there exist an linear encoder $\phi_{0} : \fR^{n} \rightarrow \fR^{k}$ and a decoder $\psi_{0} : \fR^{k} \rightarrow \fR^{n}$ such that
\begin{align*}
\Prb{ \psi_{0}\left( \phi_{0}\left( Z^{n} \right) \right) \neq Z^{n} } < \epsilon,
\end{align*}
provided that $k > \dfrac{n R_{0}}{\log \abs{\fR}}$.
Choose $\phi_{t} = \phi_{0} \circ \vec{k}_{t}$ ($t \in \S$) as the encoder for the $t$th sources and $\psi = \psi_{0} \circ \gamma$, where $\gamma : \fR^{s} \rightarrow \fR$ is defined as
$
\gamma(x_{1}, x_{2}, \cdots, x_{s}) = \sum_{t \in \S} x_{t},
$
as the decoder.
We have that
\begin{align*}
 &\Prb{ \psi\left( \phi_{1}\left( X_{1}^{n} \right),\phi_{2}\left( X_{2}^{n} \right), \cdots, \phi_{s}\left( X_{s}^{n} \right) \right) \neq Z^{n} }\\
= & \Prb{ \psi_{0}\left( \gamma\left( \phi_{0}\left( \vec{k}_{t}\left( X_{t}^{n} \right) \right) \right) \right) \neq Z^{n} }\\
= & \Prb{ \psi_{0}\left( \phi_{0}\left( \gamma\left( \vec{k}_{t}\left( X_{t}^{n} \right) \right) \right) \right) \neq Z^{n} }\\
= & \Prb{ \psi_{0}\left( \phi_{0}\left( Z^{n} \right) \right) \neq Z^{n} } < \epsilon.
\end{align*}
Therefore, $[r, r, \cdots r] \in \IR^{s}$, where $r = \dfrac{k \log
  \abs{\fR}}{n} > R_{0}$, is achievable for computing $g$. As a
conclusion, $\R \subseteq \R[g]$.  If furthermore $\fR$ is a field,
then $\fR$ is the only non-trivial left ideal of
itself. \eqref{eq:computing_c1_func} follows.
\end{IEEEproof}

The following example pictures an explicit settings of (c1) that is
not included in (c0). This example is very interesting because it
illustrates a scenario where the sources are not jointly
ergodic. Thus, \cite{Cover1975}, which assumes that the  ergodic
property holds for
the sources, does not apply. Yet, Theorem \ref{thm:computing_c1_func}
still provides a solution.

\begin{example}\label{eg:computing_Mar_func:ext}\rm
Define $\bP_{\alpha}$ and $\bP_{\beta}$ to be
\begin{align*}
 & \begin{tabular}{| c | c | c | c | c | c | c | c | c |}
\hline
 & (0, 0, 0) & (0, 0, 1) & (0, 1, 0) & (0, 1, 1) & (1, 0, 0) & (1, 0, 1) & (1, 1, 0) & (1, 1, 1)\\
 \hline
(0, 0, 0) & .1493 & .8120 & .0193 & .0193 & 0 & 0 & 0 & 0\\
\hline
(0, 0, 1) & .0193 & .9420 & .0193 & .0193 & 0 & 0 & 0 & 0\\
\hline
(0, 1, 0) & .0193 & .8120 & .1493 & .0193 & 0 & 0 & 0 & 0\\
\hline
(0, 1, 1) & .0193 & .8120 & .0193 & .1493 & 0 & 0 & 0 & 0\\
\hline
(1, 0, 0) & .0097 & .4060 & .0097 & .0097 & .1397 & .0097 & .4060 & .0097\\
\hline
(1, 0, 1) & .0097 & .4060 & .0097 & .0097 & .0097 & .1397 & .4060 & .0097\\
\hline
(1, 1, 0) & .0097 & .4060 & .0097 & .0097 & .0097 & .0097 & .5360 & .0097\\
\hline
(1, 1, 1) & .0097 & .4060 & .0097 & .0097 & .0097 & .0097 & .4060 & .1397\\
\hline
\end{tabular}\\
\mbox{and }
 & \begin{tabular}{| c | c | c | c | c | c | c | c | c |}
\hline
 & (0, 0, 0) & (0, 0, 1) & (0, 1, 0) & (0, 1, 1) & (1, 0, 0) & (1, 0, 1) & (1, 1, 0) & (1, 1, 1)\\
 \hline
(0, 0, 0) & 0 & 0 & 0 & 0 & .1493 & .8120 & .0193 & .0193\\
\hline
(0, 0, 1) & 0 & 0 & 0 & 0 & .0193 & .9420 & .0193 & .0193\\
\hline
(0, 1, 0) & 0 & 0 & 0 & 0 & .0193 & .8120 & .1493 & .0193\\
\hline
(0, 1, 1) & 0 & 0 & 0 & 0 & .0193 & .8120 & .0193 & .1493\\
\hline
(1, 0, 0) & .1493 & .8120 & .0193 & .0193 & 0 & 0 & 0 & 0\\
\hline
(1, 0, 1) & .0193 & .9420 & .0193 & .0193 & 0 & 0 & 0 & 0\\
\hline
(1, 1, 0) & .0193 & .8120 & .1493 & .0193 & 0 & 0 & 0 & 0\\
\hline
(1, 1, 1) & .0193 & .8120 & .0193 & .1493 & 0 & 0 & 0 & 0\\
\hline
\end{tabular},
\end{align*}
respectively. Let $\sM = \left\{ X^{(n)} \right\}_{-\infty}^{\infty}$ be a non-homogeneous Markov chain whose transition matrix from time $n$ to time $n+1$ is
\begin{align*}
\bP^{(n)} =
\begin{cases}
\bP_{\alpha}; & n \mbox{ is even},\\
\bP_{\beta}; & \mbox{otherwise}.
\end{cases}
\end{align*}
Consider Example \ref{eg:computing_Mar_func} by replacing the original
homogeneous Markov chain $\left\{ X^{(n)} \right\}_{-\infty}^{\infty}$
with $\sM$ defined above.  It is seen that $\sM$ does not process the
ergodic property in a strong sense \cite[pp. 68]{Hajnal1956},
i.e. $\prod_{n=1}^{\infty} \bP^{(n)}$ does not tend to a limiting
matrix with identical rows. Furthermore, there does also not exist an ``invariant
distribution'' $\pi'$ such that $\pi' \bP^{(n)} = \pi'$ for all
feasible $n$. Therefore, $\sM$ is not \emph{asymptotically mean
  stationary} \cite{Gray2009}, hence $\sM$ possesses no ergodic
property \cite[Theorem 7.1 and Theorem 8.1]{Gray2009}. As a
consequence, \cite{Cover1975} does not apply.  However, it can be
easily verified that the function $g$ is still Markovian although
$\sM$ is not even homogeneous. Moreover, it admits the same stochastic
property as shown in Example \ref{eg:computing_Mar_func}. In exact
terms, $\left\{ g\left( X^{(n)} \right) \right\}_{-\infty}^{\infty}$
is homogeneous irreducible Markovian with transition matrix $\bP$
given by \eqref{eq:1:eg_computing_Mar_func}. Consequently, Theorem
\ref{thm:computing_c1_func} offers a solution which achieves
\eqref{eq:2:eg_computing_Mar_func}.
\end{example}

For an arbitrary $g$, Lemma \ref{lma:pre_poly_func_presentation}
promises that there always exist some finite ring $\fR$ and functions
$k_{t} : \sX_{t} \rightarrow \fR$ ($t \in \S$) and $h : \fR
\rightarrow \sY$ such that
\begin{align*}
g = h\left( \sum_{t \in S} k_{t} \right).
\end{align*}
However, $k = \sum_{t \in S} k_{t}$ is not necessarily Markovian, unless the process $\sM = \left\{ X^{(n)} \right\}_{-\infty}^{\infty}$ is Markov with transition matrix $c_{1} \bU + ( 1 - c_{1} ) \mathbf{1}$ as stated in (c0). In this case, $k$ is always Markovian so claimed by Lemma \ref{lma:Markovian_func}.

\begin{corollary}\label{crl:computing_func_with_Mark_cond}
In Problem \ref{prb:2}, assume that $\left\{ X^{(n)} \right\}_{-\infty}^{\infty}$ forms an irreducible Markov chain with transition matrix $\bP_{0} = c_{1} \bU + ( 1 - c_{1} ) \mathbf{1}$, where all rows of $\bU$ are identical to some unitary vector and $0 \leq c_{1} \leq 1$. Then there exist some finite ring $\fR$ and functions $k_{t} : \sX_{t} \rightarrow \fR$ ($t \in \S$) and $h : \fR \rightarrow \sY$ such that
\begin{align}
g(x_{1}, x_{2}, \cdots, x_{s}) = h \left( \sum_{t=1}^{s} k_{t}(x_{t}) \right) \label{eq:poly_presentation}
\end{align}
and $\sM = \left\{ Z^{(n)} = \sum_{t=1}^{s} k_{t}\left( X_{t}^{(n)} \right) \right\}_{-\infty}^{\infty}$ is irreducible Markov.
Furthermore, let $\pi$ and $\bP$ be the invariant distribution and the transition matrix of $\sM$, respectively, and define
\begin{align*}
R_{0} = \max_{0 \neq \fI \leq_{l} \fR} \dfrac{\log \abs{\fR}}{\log \abs{\fI}} \min\left\{ H(\bS_{\fR / \fI} | \pi), H\left( \bP | \pi \right) - \lim_{m \rightarrow \infty} H\left( Y_{\fR / \fI}^{(m)} \left| Y_{\fR / \fI}^{(m-1)} \right. \right) \right\}
\end{align*}
where $\bS_{\fR / \fI} = \diag{ \left\{ \bS_{A} \right\}_{A \in \fR / \fI} }$ with $\bS_{A}$ being the stochastic complement of $\bP_{A, A}$ in $\bP$ and $Y_{\fR / \fI}^{(m)} = Z^{(m)} + \fI$. We have that
\begin{align}
\R_{\fR} = \left\{ [R, R, \cdots, R] \in \IR^{s} | R > R_{0} \right\} \subseteq \R[g].
\end{align}
\end{corollary}

\begin{IEEEproof}
The existences of $k_{t}$'s and $h$ are from Lemma \ref{lma:pre_poly_func_presentation}, and Lemma \ref{lma:Markovian_func} ensures that $\sM$ is Markovian. In addition, $\left\{ X^{(n)} \right\}_{-\infty}^{\infty}$ is irreducible, so is $\sM$. Finally,
$$
\lim_{m \rightarrow \infty} \dfrac{1}{m} H\left( Y_{\fR / \fI}^{(m)}, Y_{\fR / \fI}^{(m-1)}, \cdots, Y_{\fR / \fI}^{(1)} \right) = \lim_{m \rightarrow \infty} H\left( Y_{\fR / \fI}^{(m)} \left| Y_{\fR / \fI}^{(m-1)} \right. \right),
$$
since $\left\{ Y_{\fR / \fI}^{(n)} \right\}_{-\infty}^{\infty}$ is Markovian by Lemma \ref{lma:Markovian_func}. This implies that $\R_{\fR} \subseteq \R[g]$ by Theorem \ref{thm:computing_c1_func}.
\end{IEEEproof}

\begin{remark}\rm
It is easy to verify that the irreducibility requirement in (c0) is equivalent to that $u_{x} > 0$ for all $x \in \sX$.
Besides, if $c_{1} = 1$, then (c0) renders to the memoryless scenario, \cite[Problem 1]{Huang2012d}. If this is the case, Corollary \ref{crl:computing_func_with_Mark_cond} resumes corresponding results of \cite[Section VI]{Huang2012d} (see Corollary \ref{crl:computing_func_with_iid_cond}).
\end{remark}

\begin{remark}\rm
For the function $g$ in Corollary \ref{crl:computing_func_with_Mark_cond}, it is often the case that there exists more than one finite ring $\fR$ or more than one set of functions $k_{t}$'s and $h$ satisfying corresponding requirements. For example \cite{Huang2012d}, the polynomial function $x + 2 y + 3 z \in \IZ_{4}[3]$ admits also the polynomial presentation $\hat{h}\left( x + 2 y + 4 z \right) \in \IZ_{5}[3]$, where $\hat{h}(u) = \sum_{a \in \IZ_{5}} a \left[ 1 - (u-a)^{4} \right] - \left[ 1 - (u-4)^{4} \right] \in \IZ_{5}[1]$. As a conclusion, a better inner bound of $\R[g]$ is
\begin{align}
\R_{s} \bigcup \left( \bigcup_{\fR} \bigcup_{\sP_{\fR}(g)} \R_{\fR} \right), \label{eq:LC_capacity}
\end{align}
where $\sP_{\fR}(g)$ denotes all the polynomial presentations of format \eqref{eq:poly_presentation} of $g$ over ring $\fR$.
\end{remark}

\begin{corollary}\label{crl:computing_func_with_iid_cond}
In Corollary \ref{crl:computing_func_with_Mark_cond}, let $\pi = [ p_{j} ]_{j \in \fR}$. If $c_{1} = 1$, namely, $\left\{ X^{(n)} \right\}_{-\infty}^{\infty}$ and $\sM$ are i.i.d., then
\begin{align}
\R_{\fR} = \left\{ [R, R, \cdots, R] \in \IR^{s} \left| R > \max_{0 \neq \fI \leq_{l} \fR} \dfrac{\log \abs{\fR}}{\log \abs{\fI}} \left[ H(\pi) - H(\pi_{\fI}) \right] \right. \right\} \subseteq \R[g],
\end{align}
where $\pi_{\fI} = \left[ \sum_{j \in A} p_{j} \right]_{A \in \fR / \fI}$.
\end{corollary}

\begin{remark}\rm
  In Corollary \ref{crl:computing_func_with_iid_cond}, under many
  circumstances it may hold that $\max_{0 \neq \fI \leq_{l} \fR} \dfrac{\log
    \abs{\fR}}{\log \abs{\fI}} \left[ H(\pi) - H(\pi_{\fI}) \right] =
  H(\pi)$, i.e.
$$
\R_{\fR} = \left\{ [R, R, \cdots, R] \in \IR^{s} \left| R > H(\pi) \right. \right\}.
$$
For example, when $\fR$ is a field. However, $\fR$ being a field is
definitely not necessary. For more details, please kindly refer to
\cite{Huang2012d, Huang2013a, Huang2012e}.
\end{remark}

\begin{corollary}\label{crl:1:computing_func_with_Mark_cond}
In Corollary \ref{crl:computing_func_with_Mark_cond}, $\fR$ can always be chosen as a field. Consequently,
$$
\R_{\fR} = \left\{ [R, R, \cdots, R] \in \IR^{s} \left| R > H(\bP | \pi) \right. \right\} \subseteq \R[g].
$$
\end{corollary}

\begin{remark}\rm
  Although $\fR$ in Corollary \ref{crl:computing_func_with_Mark_cond}
  can always be chosen to be a field, the region $\R_{\fR}$ is not
  necessarily larger than when $\fR$ is chosen as a non-field ring. On
  the contrary, $\R_{\fR}$ is strictly larger when $\fR$ is a
  non-field ring than when it is chosen as a field in many case. This
  is because the induced $\bP$, as well as $\pi$, varies.
\end{remark}

As mentioned, in Theorem \ref{thm:computing_c1_func}, Corollary
\ref{crl:computing_func_with_Mark_cond} and Corollary
\ref{crl:computing_func_with_iid_cond}, there may be more than one
choice of such a finite ring $\fR$ satisfying the corresponding
requirements. Among those choices, $\fR$ can be either a field or a
non-field ring. Surprisingly, it is seen in (infinitely) many examples
that using non-field ring always outperforms using a field, from
several points of view. In many cases, it is proved that the
achievable region obtained with linear coding over some non-field ring
is strictly larger than any that is achieved with its field
counterpart, regardless which field is considered. \cite[Example
VI.2]{Huang2012d} has demonstrated this in the setting of correlated
i.i.d. sources. In the next section, this will be once again
demonstrated in the setting of sources with memory. In addition, other
advantages of the non-field ring linear coding technique will be
investigated in comparing with the field version.

\section{Advantages: Non-field Rings versus Fields}\label{sec:advantage}

Clearly, our discussion regarding linear coding is mainly based on
general finite rings which can be either fields or non-field rings,
each bringing their own advantages. In the setting where $g$ is the
identity function in Problem \ref{prb:2}, linear coding over finite
field is always optimal in achieving $\R[g]$ if the sources are
jointly ergodic \cite{Cover1975}.  An equivalent conclusive result is
not yet proved for linear coding over non-field ring. Nevertheless, it
is proved that there always exist more than one (up to isomorphism)
non-field rings over which linear coding achieves the Slepian--Wolf
region if the sources considered are
i.i.d. \cite{Huang2012e}. Furthermore, many examples, say Example
\ref{eg:achievability}, show that non-field ring can be equally
optimal when considering Markov sources. All in all, there is still no
conclusive support that linear coding over field is preferable in
terms of achieving the optimal region $\R[g]$ with $g$ being an
identity function.

On the contrary, there are many drawbacks of using finite fields
compared to using non-field rings (e.g. modulo integer rings):
\begin{enumerate}
\item
The finite field arithmetic is complicated to implement since the finite field arithmetic usually involves the \emph{polynomial long division algorithm}; and

\item
The alphabet size(s) of the encoder(s) is (are) usually larger than required \cite{Huang2012d, Huang2013a, Huang2013b}; and

\item
In many specific circumstances of Problem \ref{prb:2}, linear coding over any finite field is proved to be less optimal than its non-field rings counterpart in terms of achieving larger achievable region (see \cite{Huang2012d, Huang2013b} and Example \ref{eg:advantage}); and

\item The characteristic of a finite field has to be a prime. This
  constraint creates shortages in their polynomial presentations of
  discrete functions (see Lemma \ref{lma:eg}). These shortages confine
  the performance of the polynomial approach (if restrict to field)
  and lead to results like Proposition \ref{prop:eg}. On the other
  hand, The characteristic can be any positive integer for a finite
  non-field ring; and

\item
Field (finite or not) contains no \emph{zero divisor}. This also handicaps the performance of the polynomial approach (if restrict to field).

\end{enumerate}

\begin{example}\label{eg:advantage}\rm
Consider the situation illustrated in Example \ref{eg:computing_Mar_func}, one alternative is to treat that $\sX_{1} = \sX_{2} = \sX_{3} = \{ 0, 1 \}$ as a subset of finite field $\IZ_{5}$ and the function $g$ can then be presented as
\begin{align*}
g(x_{1}, x_{2}, x_{3}) = \hat{h}(x_{1} + 2 x_{2} + 4 x_{3}),
\end{align*}
where $\hat{h} : \IZ_{5} \rightarrow \IZ_{4}$ is given by
$
\hat{h}(z) =
\begin{cases}
z; & z \neq 4,\\
3; & z = 4,
\end{cases}
$
(symbol-wise).
By Corollary \ref{crl:1:computing_func_with_Mark_cond}, linear coding over $\IZ_{5}$ achieves the region
\begin{align*}
\R_{\IZ_{5}} = \left\{ [ r, r, r ] \in \IR^{3} \left| r > H\left( \bP_{\IZ_{5}} | \pi_{\IZ_{5}} \right) = 0.4623 \right. \right\}.
\end{align*}
Obviously, $\R_{\IZ_{5}} \subsetneq \R_{\IZ_{4}} \subseteq \R[g]$. In conclusion, using linear coding over field $\IZ_{5}$ is less optimal compared with over non-field ring $\IZ_{4}$. In fact, the region $\R_{\IF}$ achieved by linear coding over any finite field $\IF$ is always strictly smaller than $\R_{\IZ_{4}}$.
\end{example}

\begin{proposition}\label{prop:eg}
In Example \ref{eg:computing_Mar_func}, $\R_{\IF}$, the achievable region achieved with linear coding over any finite field $\IF$ in the sense of Corollary \ref{crl:computing_func_with_Mark_cond}, is properly contained in $\R_{\IZ_{4}}$, i.e. $\R_{\IF} \subsetneq \R_{\IZ_{4}}$.
\end{proposition}

\begin{IEEEproof}
Assume that
$$
g(x_{1}, x_{2}, x_{3}) = h\left( k_{1}(x_{1}) + k_{2}(x_{2}) + k_{3}(x_{3}) \right)
$$
with $k_{t} : \{ 0, 1 \} \rightarrow \IF$ ($1 \leq t \leq 3$) and $h : \IF \rightarrow \IZ_{4}$.
Let
\begin{align*}
\sM_{1} = \left\{ Y^{(n)} \right\}_{-\infty}^{\infty} & \mbox{ with } Y^{(n)} = g\left( X_{1}^{(n)}, X_{2}^{(n)}, X_{3}^{(n)} \right),\\
\sM_{2} = \left\{ Z^{(n)} \right\}_{-\infty}^{\infty} & \mbox{ with } Z^{(n)} = k_{1}\left( X_{1}^{(n)} \right) + k_{2}\left( X_{2}^{(n)} \right) + k_{3}\left( X_{3}^{(n)} \right),
\end{align*}
and $\bP_{l}$ and $\pi_{l}$ be the transition matrix and the invariant distribution of $\sM_{l}$, respectively, for $l = 1, 2$.
By Corollary \ref{crl:computing_func_with_Mark_cond} (also Corollary \ref{crl:1:computing_func_with_Mark_cond}), linear coding over $\IF$ achieves the region
\begin{align*}
\R_{\IF} = \left\{ [R, R, \cdots, R] \in \IR^{s} \left| R > H(\bP_{2} | \pi_{2}) \right. \right\},
\end{align*}
while linear coding over $\IZ_{4}$ achieves
\begin{align*}
\R_{\IZ_{4}} = \left\{ [R, R, \cdots, R] \in \IR^{s} \left| R > \max_{0 \neq \fI \leq_{l} \IZ_{4}} \dfrac{ \log \abs{\IZ_{4}} }{ \log \abs{\fI} } H( \bS_{\IZ_{4} / \fI} | \pi_{1} ) = H(\bP_{1} | \pi_{1}) \right. \right\}.
\end{align*}
Moreover,
\begin{align*}
H(\bP_{1} | \pi_{1}) < H(\bP_{2} | \pi_{2})
\end{align*}
by Lemma \ref{lma:data_processing_ineq} due to Lemma \ref{lma:eg} claims that $h |_{\sS}$, where $\sS = k_{1}\left( \{ 0, 1 \} \right) + k_{2}\left( \{ 0, 1 \} \right) + k_{3}\left( \{ 0, 1 \} \right)$, can never be injective.
Therefore, $\R_{\IF} \subsetneq \R_{\IZ_{4}}$.
\end{IEEEproof}

\begin{remark}\rm
  There are infinitely many functions like $g$ defined in Example
  \ref{eg:computing_Mar_func} such that the achievable region obtained
  with linear coding over any finite field in the sense of Corollary
  \ref{crl:computing_func_with_Mark_cond} is strictly suboptimal
  compared to the one achieved with linear coding over some non-field
  ring. These functions includes $\sum_{t=1}^{s} x_{t} \in \IZ_{2
    p}[s]$ for any $s \geq 2$ and any prime $p > 2$. One can always
  find a concrete example in which linear coding over $\IZ_{2 p}$
  dominates. The reason for this is partially because these functions
  are defined on rings (e.g. $\IZ_{2 p}$) of non-prime
  characteristic. However, a finite field must be of prime
  characteristic, resulting in conclusions like Proposition
  \ref{prop:eg}.
\end{remark}

As a direct consequence of Proposition \ref{prop:eg}, we have

\begin{theorem}
In the sense of \eqref{eq:LC_capacity}, linear coding over finite field is not optimal.
\end{theorem}
%
%
%
%
%
%

\section{Conclusions}

This paper considers the ring linear coding technique introduced in
\cite{Huang2012d} in the setting of compressing data generated by a
single Markov source. An achievability theorem, as a generalization of
its field counterpart, is presented. The paper also demonstrates that
the compression limit can be reached with linear encoders over
non-field rings. However, this property is not yet conclusively proved
in general.

On the other hand, a variation of the data compression problem, namely
Problem \ref{prb:2} is addressed. We apply the polynomial approach of
\cite{Huang2012a, Huang2012b, Huang2012d} to the scenarios where
sources are with memory. Once again, it is seen that linear coding
technique over non-field ring dominates its field counterpart in terms
of achieving better coding rates for computing (encoding) some
functions. On this regard, we claim that linear coding over finite
field is not optimal.

To facilitate our discussions, the concept of Supremus typical
sequence and its related asymptotic properties are introduced. These
include the AEP and four generalized typicality lemmas. The new
techniques are hopefully helpful in understanding and investigating
related problems.

%
%

\newpage

\appendices

\section{Proof of Proposition \ref{prop:aep_of_Mar_tp}}\label{app:a}

\begin{enumerate}
\item\label{itm:1:aep_of_Mar_tp}
Let $\Prb{ X^{(1)} = x^{(1)} } = c$. By definition,
\begin{align*}
 & \Prb{ \left[ X^{(1)}, X^{(2)}, \cdots, X^{(n)} \right] = \bx }\\
= & \Prb{ X^{(1)} = x^{(1)} } \prod_{i, j \in \sX} p_{i, j}^{N(i, j; \bx)}\\
= & c \exp_{2}\left[ \sum_{i, j \in \sX} N(i, j; \bx) \log p_{i, j} \right]\\
= & c \exp_{2}\left[ - n \sum_{i, j \in \sX} - \dfrac{N(i; \bx)}{n} \dfrac{N(i, j; \bx)}{N(i; \bx)} \log p_{i, j} \right]\\
= & c \exp_{2}\left[ - n \sum_{i, j \in \sX} \left( p_{i} p_{i, j} - \dfrac{N(i; \bx)}{n} \dfrac{N(i, j; \bx)}{N(i; \bx)} \right) \log p_{i, j} - p_{i} p_{i, j} \log p_{i, j} \right].
\end{align*}
In addition, there exists a small enough $\epsilon_{0} > 0$ and a $N_{0} \in \IN^{+}$ such that $\abs{ \dfrac{N(i; \bx)}{n} \dfrac{N(i, j; \bx)}{N(i; \bx)} - p_{i} p_{i, j} } < - \eta \left/ 2 \abs{\sX}^{2} \min_{i, j \in \sX} \log p_{i, j} \right.$ and $- \dfrac{ \log c }{n} < \eta / 2$ for all $\epsilon_{0} > \epsilon > 0$ and $n > N_{0}$. Consequently,
\begin{align*}
 & \Prb{ \left[ X^{(1)}, X^{(2)}, \cdots, X^{(n)} \right] = \bx }\\
> & c \exp_{2}\left[ - n \sum_{i, j \in \sX} \dfrac{ \eta }{ 2 \abs{\sX}^{2} \min_{i, j \in \sX} \log p_{i, j} } \log p_{i, j} - p_{i} p_{i, j} \log p_{i, j} \right]\\
\geq & c \exp_{2}\left[ - n \left( \dfrac{ \eta }{ 2 } - \sum_{i, j \in \sX} p_{i} p_{i, j} \log p_{i, j} \right) \right]\\
= & \exp_{2}\left[ - n \left( - \dfrac{ \log c }{n} + \dfrac{ \eta }{ 2 } + H(\bP | \pi) \right) \right]\\
> & \exp_{2}\left[ - n \left( \eta + H(\bP | \pi) \right) \right].
\end{align*}
Similarly,
\begin{align*}
 & \Prb{ \left[ X^{(1)}, X^{(2)}, \cdots, X^{(n)} \right] = \bx }\\
< & c \exp_{2}\left[ - n \sum_{i, j \in \sX} \dfrac{ - \eta }{ 2 \abs{\sX}^{2} \min_{i, j \in \sX} \log p_{i, j} } \log p_{i, j} - p_{i} p_{i, j} \log p_{i, j} \right]\\
\leq & c \exp_{2}\left[ - n \left( - \dfrac{ \eta }{ 2 } - \sum_{i, j \in \sX} p_{i} p_{i, j} \log p_{i, j} \right) \right]\\
\leq & \exp_{2}\left[ - n \left( - \dfrac{ \eta }{ 2 } + H(\bP | \pi) \right) \right]\\
< & \exp_{2}\left[ - n \left( - \eta + H(\bP | \pi) \right) \right].
\end{align*}

\item
By Boole's inequality,
\begin{align*}
\Prb{ \bX \notin \T_{\epsilon}(n, \bP) } = & \Prb{ \left( \bigcup_{i, j \in \sX} \abs{ \dfrac{N(i, j; \bX)}{N(i; \bX)} - p_{i, j} } \geq \epsilon \right) \bigcup \left( \bigcup_{i \in \sX} \abs{ \dfrac{N(i; \bX)}{n} - p_{i} } \geq \epsilon \right) }\\
\leq & \sum_{i, j \in \sX} \cPrb{ \abs{ \dfrac{N(i, j; \bX)}{N(i; \bX)} - p_{i, j} } \geq \epsilon }{ E } + \sum_{i \in \sX} \Prb{ \abs{ \dfrac{N(i; \bX)}{n} - p_{i} } \geq \epsilon },
\end{align*}
where $E = \bigcap_{i \in \sX} \left\{ \abs{ \dfrac{N(i; \bX)}{n} - p_{i} } < \epsilon \right\}$ for all feasible $i$.

By the Ergodic Theorem of Markov chains \cite[Theorem
1.10.2]{Norris1998}, $\Prb{ \abs{ \dfrac{N(i; \bX)}{n} - p_{i} } \geq
  \epsilon } \rightarrow 0$ as $n \rightarrow \infty$ for any
$\epsilon > 0$. Thus, there is an integer $N'_{0}$, such that for all $n >
N'_{0}$, $\Prb{ \abs{ \dfrac{N(i; \bX)}{n} - p_{i} } \geq \epsilon } <
\dfrac{ \eta }{ 2 \abs{\sX} }$.  On the other hand, for $\min_{i \in
  \sX} p_{i}/2 > \epsilon > 0$, $N(i; \bx) \rightarrow \infty$ as $n
\rightarrow \infty$, conditional on $E$. Therefore, by the Strong Law
of Large Numbers \cite[Theorem 1.10.1]{Norris1998}, $\cPrb{ \abs{
    \dfrac{N(i, j; \bX)}{N(i; \bX)} - p_{i, j} } \geq \epsilon }{ E }
\rightarrow 0$, $n \rightarrow \infty$. Hence, there exists $N''_{0}$,
for all $n > N''_{0}$, $\cPrb{ \abs{ \dfrac{N(i, j; \bX)}{N(i; \bX)} -
    p_{i, j} } \geq \epsilon }{ E } < \dfrac{ \eta }{ 2 \abs{\sX}^{2}
}$.  Let $N_{0} = \max\{ N'_{0}, N''_{0} \}$ and $\epsilon_{0} =
\min_{i \in \sX} p_{i} / 2 > 0$. We have $\Prb{ \bX \notin
  \T_{\epsilon}(n, \bP) } < \eta$ for all $\epsilon_{0} > \epsilon >
0$ and $n > N_{0}$.

\item
Finally, let $\epsilon_{0}$ and $N_{0}$ be defined as in \ref{itm:1:aep_of_Mar_tp}). $\abs{ \T_{\epsilon}(n, \bP) } < \exp_{2}\left[ n \left( H(\bP | \pi) + \eta \right) \right]$ follows since
\begin{align*}
1 \geq & \sum_{\bx \in \T_{\epsilon}(n, \bP)} \Prb{\bX = \bx}\\
> & \abs{ \T_{\epsilon}(n, \bP) } \exp_{2}\left[ - n \left( H(\bP | \pi) + \eta \right) \right],
\end{align*}
if $\epsilon_{0} > \epsilon > 0$ and $n > N_{0}$.
\end{enumerate}

Let $\epsilon_{0}$ be the smallest one chosen above and $N_{0}$ be the biggest one chosen. The statement is proved.

\section{Proof of Lemma \ref{lma:pre_poly_func_presentation}}\label{app:b}

Let $\IF$ be a finite field such that $\abs{\IF} \geq \abs{\sX_{t}}$ for all $1 \leq t \leq s$ and $\abs{\IF}^{s} \geq \abs{\sY}$, and let $\fR$ be the \emph{splitting field} of $\IF$ of order $\abs{\IF}^{s}$ (one example of the pair $\IF$ and $\fR$ is the $\IZ_{p}$, where $p$ is some prime, and its \emph{Galois extension} of \emph{degree} $s$). It is easily seen that $\fR$ is an $s$ dimensional vector space over $\IF$. Hence, there exist $s$ vectors $v_{1}, v_{2}, \cdots, v_{s} \in \fR$ that are linearly independent.
Let $k_{t}$ be an injection from $\sX_{t}$ to the subspace generated by vector $v_{t}$. It is easy to verify that $k = \sum_{t=1}^{s} k_{t}$ is injective since $v_{1}, v_{2}, \cdots, v_{s}$ are linearly independent. Let $k'$ be the inverse mapping of $k : \prod_{t=1}^{s} \sX_{t} \rightarrow k\left( \prod_{t=1}^{s} \sX_{t} \right)$ and $\nu : \sY \rightarrow \fR$ be any injection. We have that
$$
\hat{g} = \nu \circ g \circ k' \in \fR[s]
$$
by \cite[Lemma 7.40]{Lidl1997}.
Define $h$ to be $\nu' \circ \hat{g}$, where $\nu'$ is the inverse mapping of $\nu : \sY \rightarrow \nu\left( \sY \right)$. We have that
\begin{align*}
g = \nu' \circ \left( \nu \circ g \circ k' \right) \circ k = \nu' \circ \hat{g} \circ k = h \circ k.
\end{align*}
The statement is proved.

\begin{remark}\rm
In the proof, $k$ is chosen to be injective because the proof includes the case that $g$ is an identity function. In general, $k$ is not necessarily injective.
\end{remark}

\section{Supporting Lemmas}

\begin{lemma}\label{lma:Markovian_func}
Let $\left\{ X^{(n)} \right\}_{-\infty}^{\infty}$ be a Markov chain with countable state space $\sX$ and transition matrix $\bP_{0}$. If $\bP_{0} = c_{1} \bU + ( 1 - c_{1} ) \mathbf{1}$, where $\bU$ is a matrix all of whose rows are identical to some countably infinite unitary vector and $0 \leq c_{1} \leq 1$, then $\left\{ \Gamma\left( X^{(n)} \right) \right\}_{-\infty}^{\infty}$ is Markov for all feasible function $\Gamma$.
\end{lemma}

\begin{IEEEproof}
Let $Y^{(n)} = \Gamma\left( X^{(n)} \right)$, and assume that $[ u_{x} ]_{x \in \sX}$ is the first row of $\bU$. For any $a, b \in \Gamma\left( \sX \right)$,
\begin{align*}
 & \cPrb{ Y^{(n+1)} = b }{ Y^{(n)} = a }\\
= & \sum_{x \in \Gamma^{-1}(a)} \cPrb{ X^{(n)} = x, Y^{(n+1)} = b }{ Y^{(n)} = a }\\
= & \sum_{x \in \Gamma^{-1}(a)} \cPrb{ Y^{(n+1)} = b }{X^{(n)} = x, Y^{(n)} = a } \cPrb{ X^{(n)} = x }{ Y^{(n)} = a }\\
= & \sum_{x \in \Gamma^{-1}(a)} \cPrb{ Y^{(n+1)} = b }{X^{(n)} = x } \cPrb{ X^{(n)} = x }{ Y^{(n)} = a }\\
= &
\begin{cases}
\sum_{x \in \Gamma^{-1}(a)} \sum_{x' \in \Gamma^{-1}(b)} c_{1} u_{x'} \cPrb{ X^{(n)} = x }{ Y^{(n)} = a }; & a \neq b\\
\sum_{x \in \Gamma^{-1}(a)} \left[ 1 - c_{1} + \sum_{x' \in \Gamma^{-1}(b)} c_{1} u_{x'} \right] \cPrb{ X^{(n)} = x }{ Y^{(n)} = a }; & a = b
\end{cases}\\
= &
\begin{cases}
c_{1} \sum_{x' \in \Gamma^{-1}(b)} u_{x'} \sum_{x \in \Gamma^{-1}(a)} \cPrb{ X^{(n)} = x }{ Y^{(n)} = a }; & a \neq b\\
\left[ 1 - c_{1} + c_{1} \sum_{x' \in \Gamma^{-1}(b)} u_{x'} \right] \sum_{x \in \Gamma^{-1}(a)} \cPrb{ X^{(n)} = x }{ Y^{(n)} = a }; & a = b\\
\end{cases}\\
= &
\begin{cases}
c_{1} \sum_{x' \in \Gamma^{-1}(b)} u_{x'}; & a \neq b\\
1 - c_{1} + c_{1} \sum_{x' \in \Gamma^{-1}(b)} u_{x'}; & a = b\\
\end{cases}\\
= & \sum_{x' \in \Gamma^{-1}(b)} \cPrb{ X^{(n+1)} = x' }{ X^{(n)} = x } \left( \forall \; x \in \Gamma^{-1}(a) \right)\\
= & \sum_{x' \in \Gamma^{-1}(b)} \cPrb{ X^{(n+1)} = x' }{ X^{(n)} = x } \cPrb{ Y^{(n)} = a }{ Y^{(n)} = a, Y^{(n-1)}, \cdots } \left( \forall \; x \in \Gamma^{-1}(a) \right)\\
= & \sum_{x \in \Gamma^{-1}(a)} \sum_{x' \in \Gamma^{-1}(b)} \cPrb{ X^{(n+1)} = x' }{ X^{(n)} = x, Y^{(n)} = a, Y^{(n-1)}, \cdots }\\
 & \cPrb{ X^{(n)} = x }{ Y^{(n)} = a, Y^{(n-1)}, \cdots }\\
= & \sum_{x \in \Gamma^{-1}(a)} \sum_{x' \in \Gamma^{-1}(b)} \cPrb{ X^{(n+1)} = x', X^{(n)} = x }{ Y^{(n)} = a, Y^{(n-1)}, \cdots }\\
= & \cPrb{ Y^{(n+1)} = b }{ Y^{(n)} = a, Y^{(n-1)}, \cdots }
\end{align*}
Therefore, $\left\{ \Gamma\left( X^{(n)} \right) \right\}_{-\infty}^{\infty}$ is Markov.
\end{IEEEproof}

\begin{remark}\rm
  Lemma \ref{lma:Markovian_func} is enlightened by \cite[Theorem
  3]{Burke1958}. However, $\left\{ X^{(n)}
  \right\}_{-\infty}^{\infty}$ in this lemma is not necessary
  stationary or finite-state.
\end{remark}

\begin{lemma}\label{lma:data_processing_ineq}
Let $\sZ$ be a countable set, $\pi = [ p(z) ]_{z \in \sZ}$ and $\bP = [ p(z_{1}, z_{2}) ]_{z_{1}, z_{2} \in \sZ}$ be a non-negative unitary vector and a stochastic matrix, respectively. For any function $h : \sZ \rightarrow \sY$, if for all $y_{1}, y_{2} \in \sY$
\begin{align}
\dfrac{ p(z_{1}, y_{2}) }{p(z_{1})} = c_{y_{1}, y_{2}}, \forall \; z_{1} \in h^{-1}(y_{1}), \label{eq:1:data_processing_ineq}
\end{align}
where $c_{y_{1}, y_{2}}$ is a constant, then
\begin{align}
H\left( h\left(Z^{(2)}\right) \left| h\left(Z^{(1)}\right) \right. \right) \leq H(\bP | \pi), \label{eq:2:data_processing_ineq}
\end{align}
where $\left( Z^{(1)}, Z^{(2)} \right) \sim \pi \bP$.
Moreover, \eqref{eq:2:data_processing_ineq} holds with equality if and only if
\begin{align}
p(z_{1}, h(z_{2})) = p(z_{1}, z_{2}), \forall \; z_{1}, z_{2} \in \sZ \mbox{ with } p(z_{1}, z_{2}) > 0. \label{eq:3:data_processing_ineq}
\end{align}
\end{lemma}

\begin{IEEEproof}
By definition,
\begin{align*}
 & H\left( h\left(Z^{(2)}\right) \left| h\left(Z^{(1)}\right) \right. \right)\\
= & - \sum_{y_{1}, y_{2} \in \sY} p(y_{1}, y_{2}) \log \dfrac{ p(y_{1}, y_{2}) }{ p(y_{1}) }\\
= & - \sum_{y_{1}, y_{2} \in \sY} \sum_{z_{1} \in h^{-1}(y_{1})} p(z_{1}, y_{2}) \log \left( \left. \sum_{z_{1}' \in h^{-1}(y_{1})} p(z_{1}', y_{2}) \right/ \sum_{z_{1}'' \in h^{-1}(y_{1})} p(z_{1}'') \right)\\
\overset{(a)}{=} & - \sum_{y_{1}, y_{2} \in \sY} \sum_{z_{1} \in h^{-1}(y_{1})} p(z_{1}, y_{2}) \log \dfrac{ p(z_{1}, y_{2}) }{ p(z_{1}) }\\
= & - \sum_{y_{1}, y_{2} \in \sY} \sum_{\substack{z_{2} \in h^{-1}(y_{2}),\\ z_{1} \in h^{-1}(y_{1})}} p(z_{1}, z_{2}) \log \dfrac{ \sum_{z_{2}' \in h^{-1}(y_{2})} p(z_{1}, z_{2}') }{ p(z_{1}) }\\
\overset{(b)}{\leq} & - \sum_{y_{1}, y_{2} \in \sY} \sum_{\substack{z_{2} \in h^{-1}(y_{2}),\\ z_{1} \in h^{-1}(y_{1})}} p(z_{1}, z_{2}) \log \dfrac{ p(z_{1}, z_{2}) }{ p(z_{1}) }\\
= & - \sum_{z_{1}, z_{2} \in \sZ} p(z_{1}, z_{2}) \log \dfrac{ p(z_{1}, z_{2}) }{ p(z_{1}) }\\
= & H(\bP | \pi),
\end{align*}
where (a) is from \eqref{eq:1:data_processing_ineq}. In addition, equality holds, i.e. (b) holds with equality, if and only if \eqref{eq:3:data_processing_ineq} is satisfied.
\end{IEEEproof}

\begin{remark}\rm
$\bP$ in the above lemma can be interpreted as the transition matrix of some Markov process. However, $\pi$ is not necessary the corresponding invariant distribution. It is also not necessary that such a Markov process is irreducible.
In the meantime, \eqref{eq:2:data_processing_ineq} can be seen as a ``data processing inequality''. In addition, \eqref{eq:1:data_processing_ineq} is sufficient but not necessary for \eqref{eq:2:data_processing_ineq}, even though it is sufficient and necessary for (a) in the above proof.
\end{remark}

\begin{lemma}\label{lma:eg}
For $g$ given by \eqref{eq:1:example} and any finite field $\IF$, if there exist functions $k_{t} : \{ 0, 1 \} \rightarrow \IF$ and $h : \IF \rightarrow \IZ_{4}$, such that
\begin{align*}
g(x_{1}, x_{2}, \cdots, x_{s}) = h\left( \sum_{t=1}^{s} k_{t}(x_{t}) \right),
\end{align*}
then $h |_{\sS}$, where $\sS = k_{1}\left( \{ 0, 1 \} \right) + k_{2}\left( \{ 0, 1 \} \right) + k_{3}\left( \{ 0, 1 \} \right)$, is not injective.
\end{lemma}

\begin{IEEEproof}
Suppose otherwise, i.e. $h |_{\sS}$ is injective. Let $h' : h\left( \sS \right) \rightarrow \sS$ be the inverse mapping of $h : \sS \rightarrow h\left( \sS \right)$. Obviously, $h'$ is bijective.
By \eqref{eq:1:example}, we have
\begin{align*}
 & h'\left[ g(1, 0, 0) \right] = k_{1}(1) + k_{2}(0) + k_{3}(0)\\
= & h'\left[ g(0, 1, 1) \right] = k_{1}(0) + k_{2}(1) + k_{3}(1)\\
\neq & h'\left[ g(1, 1, 0) \right] = k_{1}(1) + k_{2}(1) + k_{3}(0)\\
= & h'\left[ g(0, 0, 1) \right] = k_{1}(0) + k_{2}(0) + k_{3}(1).
\end{align*}
Let $\tau = h'\left[ g(1, 0, 0) \right] - h'\left[ g(1, 1, 0) \right] = h'\left[ g(0, 1, 1) \right] - h'\left[ g(0, 0, 1) \right] \in \IF$. We have that
\begin{align}
 & \tau = k_{2}(0) - k_{2}(1) = k_{2}(1) - k_{2}(0) = - \tau \nonumber\\
\implies & \tau + \tau = 0. \label{eq:1:prop:Z_4_dominates_F_q}
\end{align}
\eqref{eq:1:prop:Z_4_dominates_F_q} implies that either $\tau = 0$ or $\textup{Char}(\IF) = 2$ by \cite[Proposition II.6]{Huang2012d}.
Noticeable that $k_{2}(0) \neq k_{2}(1)$, i.e. $\tau \neq 0$, by the definition of $g$.
Thus, $\textup{Char}(\IF) = 2$. Let $\rho = k_{3}(0) - k_{3}(1)$. Obviously, $\rho \neq 0$ by the definition of $g$, and $\rho + \rho = 0$ since $\textup{Char}(\IF) = 2$. Consequently,
\begin{align*}
h'\left[ g(0, 0, 0) \right]
= & k_{1}(0) + k_{2}(0) + k_{3}(0)\\
= & k_{1}(0) + k_{2}(0) + k_{3}(1) + \rho\\
= & h'\left[ g(0, 0, 1) \right] + \rho\\
= & h'\left[ g(1, 1, 0) \right] + \rho\\
= & k_{1}(1) + k_{2}(1) + k_{3}(0) + \rho\\
= & k_{1}(1) + k_{2}(1) + k_{3}(1) + \rho + \rho\\
= & h'\left[ g(1, 1, 1) \right].
\end{align*}
Therefore, $g(0, 0, 0) = g(1, 1, 1)$ since $h'$ is bijective. This is absurd!
\end{IEEEproof}

\section{Typicality Lemmas of Supremus Typical Sequences}\label{app:d}

Given a set $\sX$, a \emph{partition} $\coprod_{k \in \sK} A_{k}$ of $\sX$ is a disjoint union of $\sX$, i.e. $A_{k'} \cap A_{k''} \neq \emptyset \Leftrightarrow k' = k''$, $\bigcup_{k \in \sK} A_{k} = \sX$ and $A_{k}$'s are not empty. Obviously, $\coprod_{A \in \fR / \fI} A$ is a partition of a ring $\fR$ given the left (right) ideal $\fI$.

\begin{lemma}\label{lma:1:Sup_tp}
Given an irreducible Markov chain $\sM = \left\{ X^{(n)} \right\}_{-\infty}^{\infty}$ with finite state space $\sX$, transition matrix $\bP$ and invariant distribution $\pi = \left[ p_{j} \right]_{j \in \sX}$. Let $\coprod_{k=1}^{m} A_{k}$ be any partition of $\sX$.
For any $\eta > 0$, there exist $\epsilon_{0} > 0$ and $N_{0} \in \IN^{+}$, such that, $\forall \; \epsilon_{0} > \epsilon > 0$, $\forall \; n > N_{0}$ and $\forall \; \bx = \left[ x^{(1)}, x^{(2)}, \cdots, x^{(n)} \right] \in \S_{\epsilon}(n, \bP)$,
\begin{align*}
\abs{ S_{\epsilon}(\bx) } < & \exp_{2}\left\{ n \left[ \sum_{k=1}^{m} \sum_{j \in A_{k}} p_{j} H(\bS_{k} | \pi_{k}) + \eta \right] \right\}\\
= & \exp_{2}\left\{ n \left[ H (\bS | \pi) + \eta \right] \right\}
\end{align*}
where
\begin{align*}
S_{\epsilon}(\bx) = \left\{ \left. \left[ y^{(1)}, y^{(2)}, \cdots, y^{(n)} \right] \in \S_{\epsilon}(n, \bP) \right| y^{(l)} \in A_{k} \Leftrightarrow x^{(l)} \in A_{k}, \forall \; 1 \leq l \leq n, \forall \; 1 \leq k \leq m \right\},
\end{align*}
$\bS_{k}$ is the stochastic complement of $\bP_{A_{k}, A_{k}}$ in $\bP$, $\pi_{k} = \dfrac{ \left[ p_{i} \right]_{i \in A_{k}} }{\sum_{j \in A_{k}} p_{j}}$ is the invariant distribution of $\bS_{k}$ and
$$
\bS = \diag{ \left\{ \bS_{k} \right\}_{1 \leq k \leq m} }.
$$
\end{lemma}

\begin{IEEEproof}
Let
$$
\bx_{A_{k}} = \left[ x^{(n_{1})}, x^{(n_{2})}, x^{(n_{m_{k}})} \right]
$$
be the subsequence of $\bx$ formed by all those $x^{(l)}$'s belong to $A_{k}$ in the original ordering. Obviously, $\sum_{k=1}^{m} m_{k} = n$ and $\abs{ \dfrac{m_{k}}{n} - \sum_{j \in A_{k}} p_{j} } < \abs{A_{k}} \epsilon + \dfrac{1}{n}$.
For any $\by = \left[ y^{(1)}, y^{(2)}, \cdots, y^{(n)} \right] \in S_{\epsilon}(\bx)$, it is easily seen that
$$
\by_{A_{k}} = \left[ y^{(n_{1})}, y^{(n_{2})}, y^{(n_{m_{k}})} \right] \in A_{k}^{m_{k}}
$$
is a strongly Markov $\epsilon$-typical sequence of length $m_{k}$ with respect to $\bS_{k}$, since $\by$ is Supremus $\epsilon$-typical. Additionally, by Proposition \ref{prop:aep_of_Mar_tp}, there exist $\epsilon_{k} > 0$ and positive integer $M_{k}$ such that the number of strongly Markov $\epsilon$-typical sequences of length $m_{k}$ is upper bounded by $\exp_{2}\left\{ m_{k} \left[ H(\bS_{k} | \pi_{k}) + \eta/2 \right] \right\}$ if $0 < \epsilon < \epsilon_{k}$ and $m_{k} > M_{k}$. Therefore, if $0 < \epsilon < \min_{1 \leq k \leq m} \epsilon_{k}$, $n > M = \max_{1 \leq k \leq m}\left\{ \dfrac{ 1 + M_{k} }{ \abs{\sum_{j \in A_{k}} p_{j} - \abs{A_{k}} \epsilon} } \right\}$ (this guarantees that $m_{k} > M_{k}$ for all $1 \leq k \leq m$), then
\begin{align*}
\abs{S_{\epsilon}(\bx)} \leq & \exp_{2}\left\{ \sum_{k=1}^{m} m_{k} \left[ H(\bS_{k} | \pi_{k}) + \eta/2 \right] \right\}\\
= & \exp_{2}\left\{ n \left[ \sum_{k=1}^{m} \dfrac{m_{k}}{n} H(\bS_{k} | \pi_{k}) + \eta/2 \right] \right\}.
\end{align*}
Furthermore, choose $0 < \epsilon_{0} \leq \min_{1 \leq k \leq m} \epsilon_{k}$ and $N_{0} \geq M$ such that $\dfrac{m_{k}}{n} < \sum_{j \in A_{k}} p_{j} + \dfrac{\eta}{2 \sum_{k=1}^{m} H(\bS_{k} | \pi_{k})}$ for all $0 < \epsilon < \epsilon_{0}$ and $n > N_{0}$ and $1 \leq k \leq m$, we have
\begin{align*}
\abs{S_{\epsilon}(\bx)} < & \exp_{2}\left\{ n \left[ \sum_{k=1}^{m} \sum_{j \in A_{k}} p_{j} H(\bS_{k} | \pi_{k}) + \eta \right] \right\},
\end{align*}
\eqref{ineq:1:diff_typical_set} is established. Direct calculation yields \eqref{ineq:2:diff_typical_set}.
\end{IEEEproof}

\begin{lemma}\label{lma:2:Sup_tp}
In Lemma \ref{lma:1:Sup_tp}, define $\Gamma(x) = l \Leftrightarrow x \in A_{l}$. We have that
\begin{align*}
\abs{ S_{\epsilon}(\bx) } < \exp_{2}\left\{ n \left[ H (\bP | \pi) - \lim_{w \rightarrow \infty} \dfrac{1}{w} H\left( Y^{(w)}, Y^{(w-1)}, \cdots, Y^{(1)} \right) + \eta \right] \right\},
\end{align*}
where $Y^{(w)} = \Gamma\left( X^{(w)} \right)$.
\end{lemma}

\begin{IEEEproof}
Let
$$
\overline{\by} = \left[ \Gamma\left( x^{(1)} \right), \Gamma\left( x^{(2)} \right), \cdots, \Gamma\left( x^{(n)} \right) \right].
$$
By definition,
$$
\left[ \Gamma\left( y^{(1)} \right), \Gamma\left( y^{(2)} \right), \cdots, \Gamma\left( y^{(n)} \right) \right] = \overline{\by},
$$
for any $\by = \left[ y^{(1)}, y^{(2)}, \cdots, y^{(n)} \right] \in S_{\epsilon}(\bx)$.
$\by$ is \emph{jointly typical} \cite{Cover1975} with $\overline{\by}$ with respect to the process
\begin{align*}
\cdots,
\left(\begin{matrix}
X^{(1)}\\
Y^{(1)}
\end{matrix}\right),
\left(\begin{matrix}
X^{(2)}\\
Y^{(2)}
\end{matrix}\right),
\cdots,
\left(\begin{matrix}
X^{(n)}\\
Y^{(n)}
\end{matrix}\right),
\cdots
\end{align*}
Therefore, there exist $\epsilon_{0} > 0$ and $N_{0} \in \IN^{+}$, such that, $\forall \; \epsilon_{0} > \epsilon > 0$ and $\forall \; n > N_{0}$,
\begin{align*}
\abs{ S_{\epsilon}(\bx) } < & \exp_{2}\bigg\{ n \bigg[ \lim_{w \rightarrow \infty} \dfrac{1}{w} H\left( X^{(w)}, X^{(w-1)}, \cdots, X^{(1)} \right)\\
 & - \lim_{w \rightarrow \infty} \dfrac{1}{w} H\left( Y^{(w)}, Y^{(w-1)}, \cdots, Y^{(1)} \right) + \eta \bigg] \bigg\}\\
= & \exp_{2}\bigg\{ n \bigg[ H\left( \bP | \pi \right) - \lim_{w \rightarrow \infty} \dfrac{1}{w} H\left( Y^{(w)}, Y^{(w-1)}, \cdots, Y^{(1)} \right) + \eta \bigg] \bigg\},
\end{align*}
where the equality follows from the fact that $\lim_{w \rightarrow \infty} \dfrac{1}{w} H\left( X^{(w)}, X^{(w-1)}, \cdots, X^{(1)} \right) = H\left( \bP \left| \pi \right. \right)$ since $\sM$ is irreducible Markov.
\end{IEEEproof}

%
%

\ifCLASSOPTIONcaptionsoff
  \newpage
\fi

\bibliography{Ref.bib}
\bibliographystyle{IEEEtran}

%
%





\end{document}